\documentclass[aps,pre,reprint,superscriptaddress,showpacs,notitlepage,a4]{revtex4-2}
\usepackage{amssymb,amsmath}

\usepackage{lineno,hyperref}
\usepackage{graphicx,epstopdf}
\usepackage{epsfig}
\usepackage{subfigure}
\usepackage{amsmath,amssymb,amstext}
\usepackage[export]{adjustbox}
\usepackage{float}
\usepackage{comment}
\usepackage{xcolor}
\usepackage{appendix}
\usepackage{dcolumn}% Align table columns on decimal point
\usepackage{bm}% bold mathcapabilities
\usepackage[normalem]{ulem}

\newcommand{\rhop}{\rho^{+}}
\newcommand{\rhom}{\rho^{-}}
\newcommand{\rhos}{\rho^\mathrm{s}}
\newcommand{\rhou}{\rho^\mathrm{u}}
\newcommand{\nb}{\mathbf{n}}
\newcommand{\st}{\mathrm{st}}

\begin{document}

\title{Partisan voter model on complex networks: Dynamics of local ordering}% 
\author{Jaume Llabr\'es}
\email{jaumellabres@ifisc.uib-csic.es}
 \affiliation{Instituto de Fisica Interdisciplinar y Sistemas Complejos IFISC (CSIC-UIB), Campus UIB, 07122 Palma de Mallorca, Spain}
\author{Maxi San Miguel}
\email{maxi@ifisc.uib-csic.es}
\affiliation{Instituto de Fisica Interdisciplinar y Sistemas Complejos IFISC (CSIC-UIB), Campus UIB, 07122 Palma de Mallorca, Spain}
\author{Ra\'ul Toral}
\email{raul@ifisc.uib-csic.es}
\affiliation{Instituto de Fisica Interdisciplinar y Sistemas Complejos IFISC (CSIC-UIB), Campus UIB, 07122 Palma de Mallorca, Spain}

\date{\today}

\begin{abstract}
We investigate the processes of local ordering for the partisan voter model on complex networks. In this model, agents hold a binary opinion and a fixed preference that biases updates toward alignment with their preferred opinion.
We first study the dynamics on uncorrelated random networks and derive a pair approximation that resolves the densities of links connecting different classes of agents. The analytical predictions are in excellent agreement with Monte Carlo simulations. In this setting, partisan bias leaves the total stationary density of links connecting nodes in different sates unchanged and at the same value as in the standard voter model, but redistributes it among different categories of links.
We then consider preference-dependent networks with homophilic and heterophilic attachment to analyze the competition between the global bias mechanism and the local effect of preference-based connectivity. In this case, structural correlations qualitatively modify the stationary state. We identify different regimes of local ordering in the space of parameters measuring the strength of the preference and the strength of the homophilic attachment.
Our work clarifies the distinct roles of dynamical partisan bias and structural assortativity, and provides an analytical framework to study partisan opinion dynamics beyond mean-field theory.
\end{abstract}

\maketitle

\section{Introduction}
% Intro voter model
Understanding how collective opinion emerges from local social interactions is a central problem in the statistical physics of social dynamics~\cite{Castellano_2009, Starnini_2025}. Among the many models proposed to address this question, the \emph{voter model} occupies a distinguished place because of its simplicity and analytical tractability~\cite{Clifford_1973,Holley_1975}. In its standard form, agents hold one of two possible opinions and update their state by copying that of a randomly chosen neighbor. Despite the simplicity of this microscopic rule, the model exhibits fluctuation-driven ordering, absorbing consensus states, nontrivial finite-size effects, and a strong sensitivity to dimensionality, disorder, and interaction topology~\cite{Sood_2005,Suchecki_2005,Sood_2008,Vazquez_2008}. 
A key property of the voter model is the conservation of the average magnetization, which makes consensus a fluctuation-driven outcome rather than the deterministic relaxation toward a preferred state.
Beyond its role as a theoretical benchmark, the voter model has also been successfully used to describe data from electoral processes~\cite{FernandezGracia_2014}. It has also served as a reference point for studying universality classes and phase transitions in symmetric binary-state systems~\cite{Hammal_2005,Llabres_2026}.

% Bias in the voter model 
A key limitation of the standard voter model is its dynamical neutrality: agents have no intrinsic preference for either state. This assumption is often too restrictive in social and political contexts, where individual responses to social influence may be biased by persistent predispositions, identities, or partisan preferences. Several extensions of voter-like dynamics have therefore incorporated quenched heterogeneity, asymmetric influence, or internal biases~\cite{Mobilia_2003, Masuda_2010, Vazquez_2010, Masuda_2011, Redner_2019, Czaplicka_2022}. In this context, the partisan voter model assigns to each agent a fixed binary preference that biases opinion changes toward alignment with that preference~\cite{Masuda_2010, Masuda_2011, Llabres_2023}. From a statistical-physics viewpoint, this preference acts as a quenched internal field that breaks the conservation of magnetization of the standard voter model. Previous work has shown that this symmetry-breaking perturbation induces a stable \emph{coexistence} of both states in the mean-field limit~\cite{Masuda_2010,Llabres_2023}. The partisan voter model therefore provides a minimal framework to study the competition between social imitation and individual predisposition.

% Network effects VM
The role of network structure has been extensively studied for the standard voter model. On complex networks, finite connectivity and degree heterogeneity modify the ordering process, the density of \emph{active links}, i.e. links connecting nodes in opposite states, and the time needed to reach one of the two absorbing states~\cite{ Suchecki_2005, Sood_2005, Sood_2008,Vazquez_2008, Suchecki_Euro_2005, Castellano_2005}.
Many of these effects can be captured analytically by pair approximations, which go beyond mean-field theory by retaining the density of active links as a dynamical variable~\cite{Vazquez_2008}. More generally, pair approximations and related moment-closure schemes have become useful tools for binary-state dynamics on networks, providing analytically tractable descriptions of local correlations~\cite{Gleeson_2011,Kuehn_2016, Peralta_2018, Peralta_Sto_2018, Ramirez_2024}

% Proposal of the PA for the PVM
Extending this type of description to the partisan voter model is necessary to understand how fixed preferences interact with sparse and heterogeneous connectivity beyond the fully connected limit. In the standard voter model, all active links are dynamically equivalent, so a single density of active links characterizes local disorder at the pair level. In the partisan voter model, however, agents hold both a dynamical state and a fixed preference. As a result, active links are no longer equivalent: they must be distinguished according to the preferences of the connected agents. This decomposition of the total density of active links is a central analytical ingredient of the present work.

% Homophilic networks
A particularly relevant source of structural correlations in social systems is homophily, namely, the tendency of individuals to connect preferentially with others who are similar to them~\cite{McPherson_2001}. Homophily is known to promote segregation and polarization in models of social influence~\cite{Flache_2011,Mas_2013}, and to affect visibility, perception biases, and information transmission in social networks~\cite{Karimi_2018,Karimi_2019,diaz_homophily}. In the present context, homophily implies that the same fixed preferences that bias individual updates may also organize the network of social contacts. The preference for a given state thus plays a dual role: it acts dynamically, by modifying the transition probabilities, and structurally, by inducing assortative or disassortative mixing. The interplay between these two mechanisms is expected to affect both the global state of coexistence of the two states and the local ordering of the system, which is measured by the density of active links.

% Dynamical role
In this work, we study the partisan voter model on complex networks by separating the dynamical and structural roles associated with fixed preferences. We first consider uncorrelated random networks, where preferences bias the update rule but do not affect the structure of interactions. The strength of this bias is quantified by the parameter $\varepsilon$~\cite{Llabres_2023}, defined as the difference between the probabilities of adopting the preferred and the non-preferred opinions. For this case, we develop a pair approximation for the different classes of active links defined by the preferences of the connected agents and provide an analytical description beyond mean-field theory. The preference strength acts as a dynamical perturbation: it breaks the conservation of magnetization of the standard voter model and induces a coexistence state. However, this perturbation does not change the total stationary density of active links. Instead, it reorganizes local order by redistributing active links among their different classes.

% Structural role
We then study networks with preference-based attachment, where the same preferences also shape the structure of interactions. The tendency to form links between agents with the same preference is quantified by the homophily parameter $h$~\cite{Karimi_2018}, which introduces preference correlations in the network. For $h>1/2$, links are formed preferentially between agents with the same preference, leading to homophilic networks, whereas for $h<1/2$, links are formed preferentially between agents with opposite preferences, leading to heterophilic networks. In this case, structural correlations modify both the total stationary density of active links and its decomposition among the different classes. The interplay between $\varepsilon$ and $h$ gives rise to three different regimes in which preference-based network organization can either enhance or suppress local order in the system.

Our results show that, in binary-state dynamics with quenched agent types, the total density of active links is not sufficient to characterize local order. A finer description resolving the different classes of active links is therefore required to accurately describe how activity is organized and evolves in the system.

% Paper cont
The paper is organized as follows. In Sec.~\ref{sec:model}, we define the partisan voter model on complex networks and introduce the relevant observables. In Sec.~\ref{sec:PA}, we present the pair approximation for uncorrelated random networks and assess its accuracy against numerical simulations of the model. In Sec.~\ref{sec:homophily}, we investigate how homophilic and heterophilic attachment affect the stationary behavior of the system. Finally, in Sec.~\ref{sec:conclusions}, we summarize the main results and discuss possible extensions.

\section{The partisan voter model on complex networks}\label{sec:model}
Let us consider a population of $N$ agents placed on the nodes of a static undirected network. We denote by $\nu(i)$ the set of neighbors of node $i$, and by $k_i$ its degree. The total number of links in the network is $E=\frac12\sum_ik_i$.

In the partisan voter model~\cite{Masuda_2010, Masuda_2011, Llabres_2023}, each agent $i \in~[1,N]$ is characterized by a binary state $s_i \in \{-1,+1\}$ and a binary preference $p_i \in \{-1,+1\}$. Thus, there exist two types of agents, each of them being in two possible states. The preference $p_i$ remains constant throughout the dynamics while the state $s_i$ evolves according to the following update rules: At each time step, an agent $i$ is randomly selected, and one of its neighbors $j \in \nu(i)$ is chosen uniformly at random. If $s_i = s_j$, nothing happens. Otherwise, agent $i$ updates its state with a probability that depends on its preference $p_i$
\begin{equation}
 P(s_i\to-s_i)=\frac{1-s_i p_i \varepsilon}{2},
\end{equation}
so that an agent copies its preferred opinion with probability $\tfrac12(1+\epsilon)$ and the unpreferred opinion with probability $\tfrac12(1-\epsilon)$. Time is measured in Monte Carlo steps (MCS). One MCS consists of $N$ update attempts, so that, on average, each agent is selected once per MCS.

The model presents two absorbing states, which we refer to as \textit{consensus} states, corresponding to the two configurations in which all agents are in the same state, either $+1$ or $-1$, and from which no further evolution is possible. The parameter $\varepsilon \in [0,1]$ quantifies the strength of the preference and is assumed to be homogeneous across agents. For $\varepsilon=0$, agents copy their neighbors with probability $1/2$, thus recovering the standard voter model, with an irrelevant global time rescaling. In contrast, for $\varepsilon=1$, agents whose state is already aligned with their preference become zealots, as they never change state. As a result, in this limiting case, the system eventually reaches a frozen coexistence configuration in which each agent is aligned with its own preference. 

Let $N^p_s$ denote the global number of agents of class $(s,p)$, i.e., agents with state $s$ and preference $p$, and define the corresponding densities as $x^p_s = N^p_s / N$. We assume that preferences are equally distributed, such that half of the population has $p=+1$ and the other half $p=-1$, implying that $x^+_++x^+_-=x^-_++x^-_-=1/2$. This constraint together with the conservation of the number of agents allow us to characterize the macroscopic state of the system with two independent variables. We choose the imbalance $\Delta\equiv~ x^+_+ - x^-_-$ and the total density $\Sigma \equiv x^+_+ + x^-_-$ of satisfied agents, i.e., agents $i$ whose state is aligned with their preference, $s_i=p_i$. By construction, $\Delta \in [-1/2,1/2]$ and $\Sigma \in [0,1]$. The imbalance $\Delta$ is directly related to the magnetization,
$m = \tfrac 1N \sum_{i=1}^N s_i = 2\Delta$.

A link $(i,j)$ between nodes $i$ and $j$ is said to be \emph{active} if $s_i \neq s_j$. The level of local order of the system is characterized by the density of active links,
\begin{equation}
\rho = \frac{1}{4E} \sum_{i=1}^N \sum_{j \in \nu(i)} (1 - s_i s_j).
\end{equation}
Note that this quantity vanishes in the absorbing states. 

Since the dynamics depends explicitly on the preferences of the agents, the global density of active links is not sufficient to fully characterize the local order of the system. It is therefore necessary to distinguish how active links are distributed with respect to the preferences of the connected agents. To this end, we decompose $\rho$ into four contributions:
\begin{itemize}
 \item $\rhop$: Density of active links between agents with preference $p=+1$,
 \begin{equation} \label{eq:rhop_def}
 \rhop = \frac{1}{16E} \sum_{i=1}^N \sum_{j \in \nu(i)} (1 - s_i s_j)(1+p_i)(1+p_j),
 \end{equation}

 \item $\rhom$: Density of active links between agents with preference $p=-1$,
 \begin{equation} \label{eq:rhom_def}
 \rhom = \frac{1}{16E} \sum_{i=1}^N \sum_{j \in \nu(i)} (1 - s_i s_j)(1-p_i)(1-p_j),
 \end{equation}

 \item $\rhos$: Density of active links between satisfied agents, i.e., nodes whose state is aligned with its preference
 \begin{equation} \label{eq:rhos_def}
 \rhos = \frac{1}{16E} \sum_{i=1}^N \sum_{j \in \nu(i)} (1 - s_i s_j)(1+s_i p_i)(1+s_j p_j),
 \end{equation}

 \item $\rhou$: Density of active links between unsatisfied agents, i.e., nodes whose state is unaligned with its preference
 \begin{equation} \label{eq:rhou_def}
 \rhou = \frac{1}{16E} \sum_{i=1}^N \sum_{j \in \nu(i)} (1 - s_i s_j)(1-s_i p_i)(1-s_j p_j).
 \end{equation}
\end{itemize}
When added together, these variables give us the total density of active links
\begin{equation}
\rho = \rhop + \rhom + \rhos + \rhou.
\end{equation}

The set $\{\Delta, \Sigma, \rhop, \rhom, \rhos, \rhou\}$ defines the macroscopic stochastic variables we use to characterize the dynamical behavior of the system. 

\subsection{Mean-field limit}
In the complete graph, the densities of active links can be expressed in terms of the global variables $\Delta$ and $\Sigma$
\begin{subequations}\label{eq:rhomf}
\begin{align}
\rhop &= \frac{1}{2}(\Sigma+\Delta)(1-\Sigma-\Delta), \\
\rhom &= \frac{1}{2}(\Sigma-\Delta)(1-\Sigma+\Delta), \\
\rhos &= \frac{1}{2}(\Sigma+\Delta)(\Sigma-\Delta), \\
\rhou &= \frac{1}{2}(1-\Sigma+\Delta)(1-\Sigma-\Delta).
\end{align}
\end{subequations}
Summing these contributions, the total density of active links reads
\begin{equation} \label{eq:rho_delta_MF}
\rho = \frac{1}{2} - 2\Delta^2,
\end{equation}
showing that $\rho$ and $\Delta$ are not independent variables in the complete graph. This relation is mandatory and equivalent to that obtained for the standard voter model $\rho=\tfrac12 (1-~m^2)$.

In the mean-field, i.e. complete graph and thermodynamic limit, $N\to\infty$, the dynamical evolution of the system is described by a set of deterministic equations~\cite{Masuda_2010, Llabres_2023}
\begin{align}\label{eq:mfdelta}
\frac{d\Delta}{dt} &= \varepsilon \Delta \left(1 - 2\Sigma\right), \\ \label{eq:mfsigma}
\frac{d\Sigma}{dt} &= \frac{1 + \varepsilon}{2} - \Sigma - 2\varepsilon\Delta^2.
\end{align}
This dynamical system exhibits the following fixed points:
\begin{itemize}
 \item Two absorbing, saddle fixed points, corresponding to consensus, 
 \begin{equation}
 (\Delta_\st,\Sigma_\st)=\left(\pm \frac 12, \frac 12\right),
 \end{equation}
 for which the total density of active links $\rho$ and all its contributions vanish, $\rho_\st=\rhop_\st= \rhom_\st=\rhos_\st= \rhou_\st=0$.
 \item One stable coexistence fixed point 
 \begin{equation}
 (\Delta_\st,\Sigma_\st)=\left(0, \frac{1+\varepsilon}{2}\right),
 \end{equation}
 for which the total stationary density of active links is independent of the preference, $\rho=1/2$, and the different contributions read
\begin{align} \label{eq:rhos_MF}
 &\left(\rhop_\st, \rhom_\st, \rhos_\st, \rhou_\st\right)= \nonumber \\ 
 &\left(\frac{1-\varepsilon^2}{8},\frac{1-\varepsilon^2}{8}, \frac{(1+\varepsilon)^2}{8}, \frac{(1-\varepsilon)^2}{8}\right).
\end{align}
\end{itemize}

\begin{figure*}[t]
 \centering \includegraphics[width=0.32\linewidth]{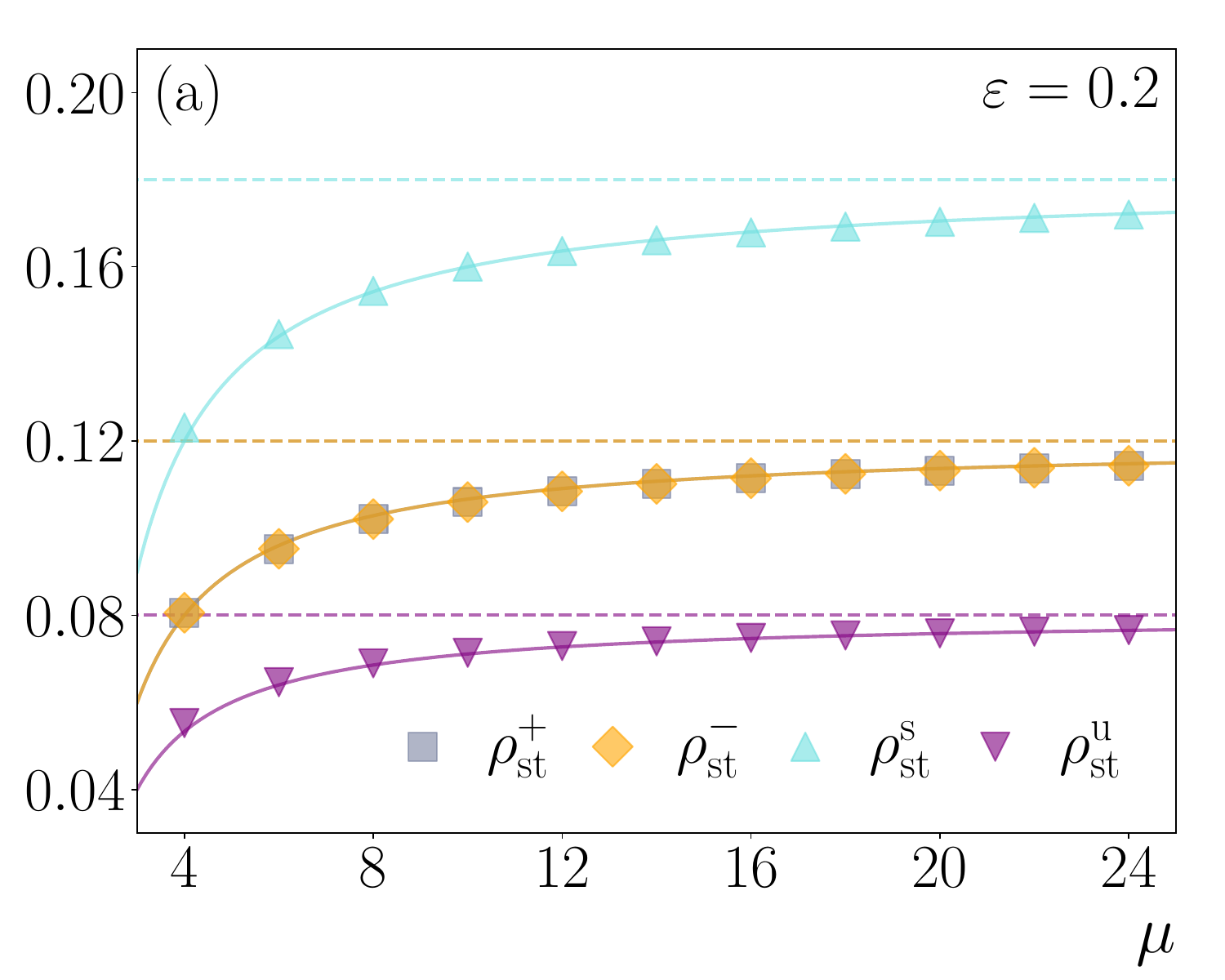}
 \includegraphics[width=0.32\linewidth]{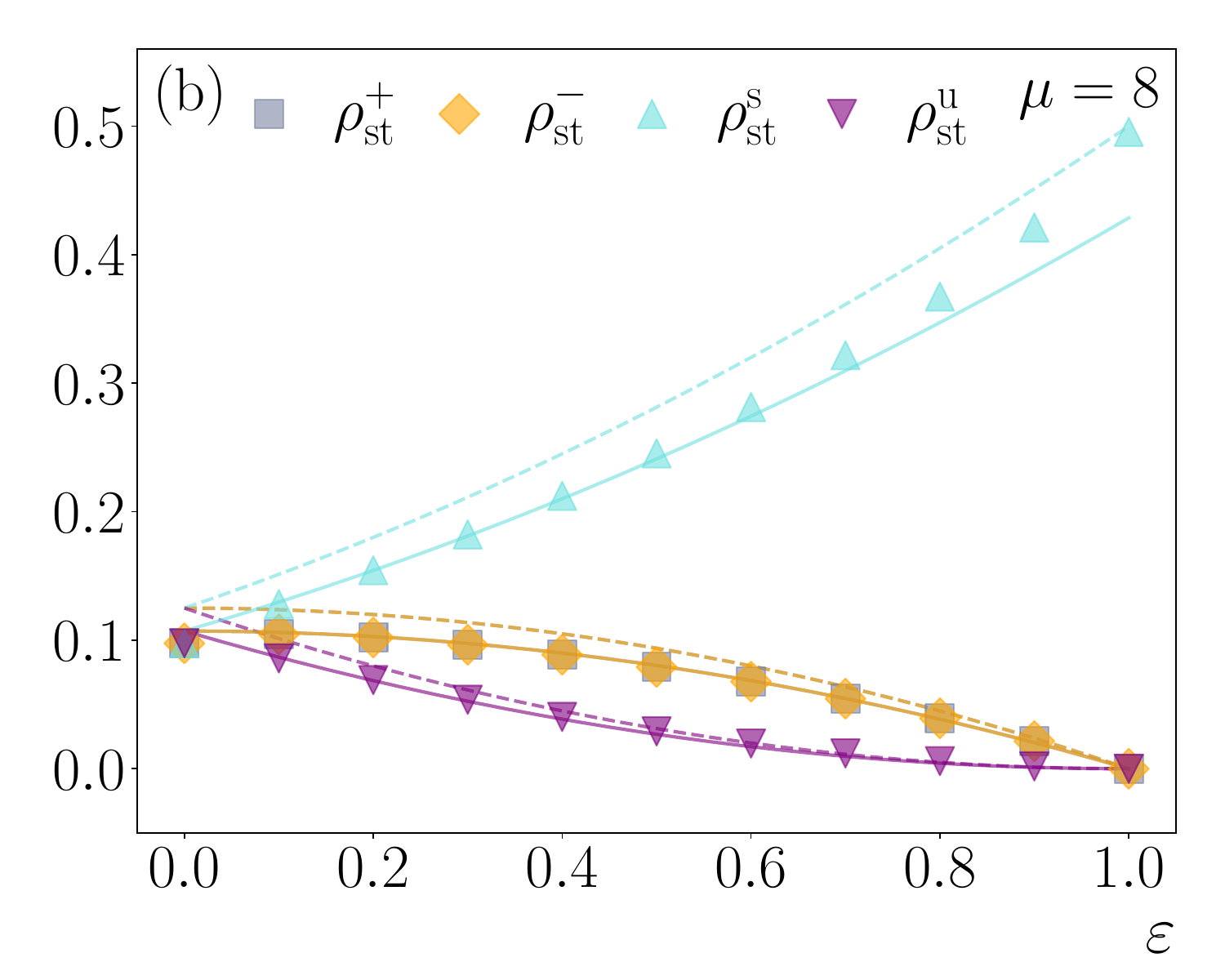}
 \includegraphics[width=0.32\linewidth]{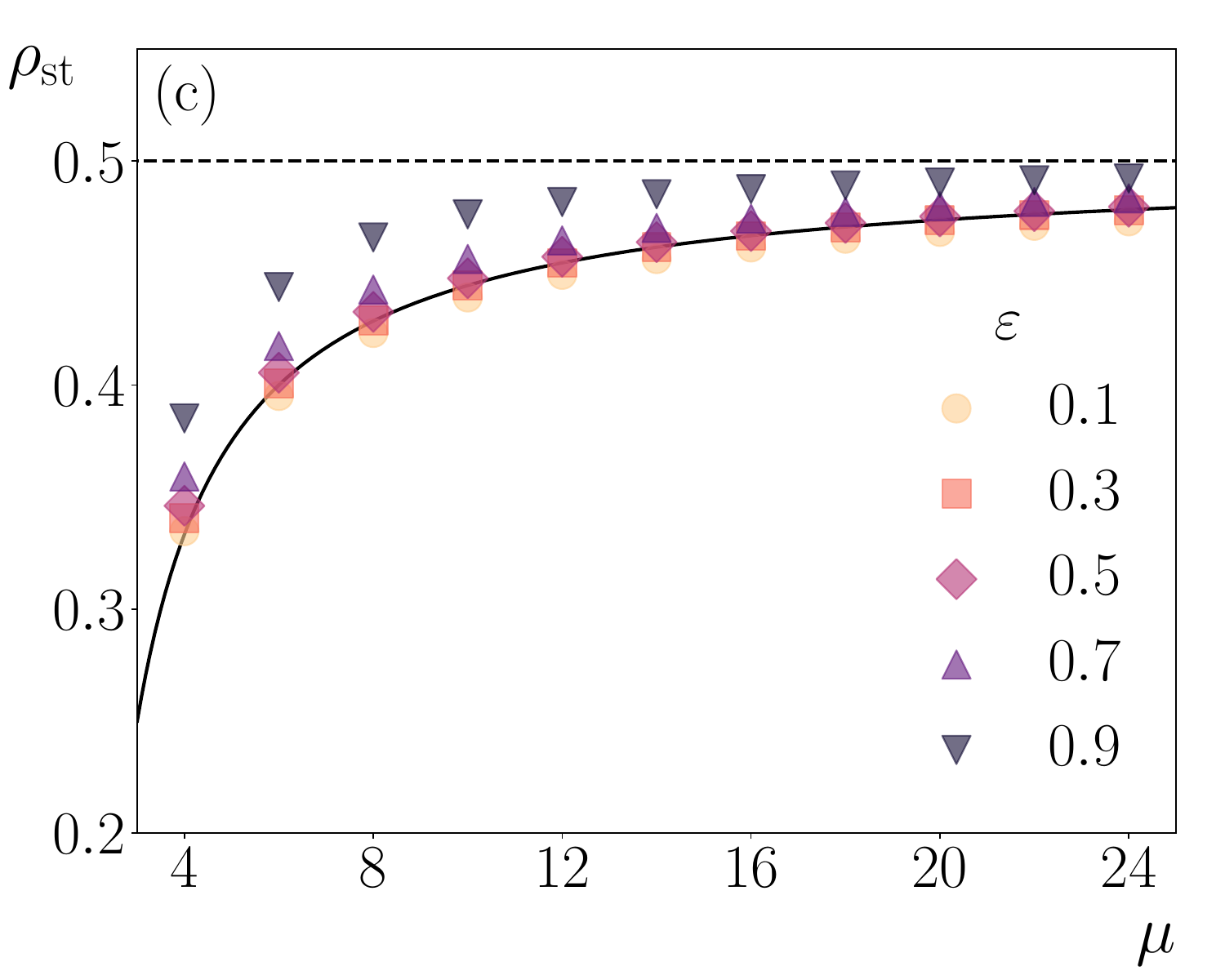}
 \caption{\textbf{Partisan voter model on uncorrelated networks. Stationary state of the densities of active links.} Stationary values of the densities of active links $\rhop_\st$, $\rhom_\st$, $\rhos_\st$, and $\rhou_\st$, corresponding, respectively, to links between nodes with preference $p=+1$, preference $p=-1$, satisfied nodes, and unsatisfied nodes, (a) as a function of the average degree $\mu$ for a value of the preference $\varepsilon=0.2$, and (b) as a function of $\varepsilon$ for $\mu=8$. Note that data for $\rhop$ and $\rhom$ overlap. (c) Stationary values of the total density of active links $\rho_\st$, as a function of $\mu$ for several values of $\varepsilon$, as indicated in the legend. Symbols correspond to numerical simulations of the agent-based model on Erd\H{o}s--R\'enyi networks of size $N=10000$, solid lines represent the pair approximation solution given by Eqs.~\eqref{eq:PA_solution}, \eqref{eq:xi}, \eqref{eq:PA_solution_rho}, and dashed lines indicate the mean-field prediction, Eqs.~\eqref{eq:rhos_MF}.}
 \label{fig:PA_ER}
\end{figure*}

While on the complete graph the dynamics can be fully described in terms of the two variables $\Delta$ and $\Sigma$, such a reduction is no longer possible in structured networks. We address this case in the following section.

\section{Pair approximation} \label{sec:PA}
In this section, we study the partisan voter model beyond the mean-field description by incorporating the effects of network structure through a pair approximation. For the standard voter model this approximation couples the evolution of the magnetization and the density of active links, and it is known to perform well for uncorrelated networks with arbitrary degree distributions~\cite{Vazquez_2008}. A technical difficulty arises for the partisan voter model because, in addition to the global variables $\Delta$ and $\Sigma$, we need to keep track of the densities of active links connecting different types of agents, $\rhop$, $\rhom$, $\rhos$, and $\rhou$, as introduced in Eqs.~\eqref{eq:rhop_def}--\eqref{eq:rhou_def}.

We extend now the homogeneous pair approximation by following closely the formalism introduced in Ref.~\cite{Vazquez_2008} for the standard voter model, and further developed for the nonlinear voter model~\cite{Ramirez_2024}. Within this framework, the time evolution of an observable $X$ is described by
\begin{align} \label{eq:HPA_general}
 \frac{dX}{dt}=&\frac{1}{\delta t}\sum_{k}\sum_{s=\pm}\sum_{p=\pm}\sum_{\nb}P(s,p,k,\nb) \nonumber \\ & \times P_{s\to-s}(s,p,k,\nb)\delta X(s\to-s,\nb),
\end{align}
where $\delta t=1/N$ is the elementary time step. Here, $P(s,p,k,\nb)$ is the probability of selecting a node with state $s$, preference $p$, degree $k$, and neighborhood composition $\nb\equiv(n_+^+,n_-^+,n_+^-,n_-^-)$, where $n_{s}^{p}$ denotes the number of neighbors of class $(s,p)$ satisfying
\begin{equation}
 n_{+}^{+}+n_{-}^{+}+n_{+}^{-}+n_{-}^{-}=k.
\end{equation}
The term $P_{s\to -s}(s,p,k,\nb)$ is the probability that the selected node flips its state
\begin{equation}\label{eq:p_flip}
P_{s\to -s}(s,p,k,\nb)
=
\frac{n_{-s}}{k}
\left(\frac{1-sp\varepsilon}{2}\right),
\end{equation}
where $ n_{-s}\equiv n^{+}_{-s}+n^{-}_{-s}$ is the number of neighbors in the opposite state. The first factor corresponds to the probability of selecting a neighbor in an opposite state, while the second encodes the bias induced by the preference. Finally, $\delta X(s\to -s,\nb)$ is the corresponding variation in the observable $X$ when the flip occurs.

The probability $P(s,p,k,\nb)$ is assumed to factorize as
\begin{equation} \label{eq:P_spnk}
P(s,p,k,\nb)=x_s^p\,P_k\,M_s^p(k,\nb),
\end{equation}
where $P_k$ is the degree distribution and $M_s^p(k,\nb)$ denotes the probability that a node of class $(s,p)$ has neighborhood composition $\nb$. Under the pair approximation, we further assume that $M_s^p(k,\nb)$ is well described by a multinomial distribution,
\begin{align}
M_s^p(k,\nb)
&=
\frac{k!}{n_+^+!\,n_-^+!\,n_+^-!\,n_-^-!}
\prod_{s'=\pm}\prod_{p'=\pm}
\left[P(s',p'|s,p)\right]^{n_{s'}^{p'}},
\label{eq:multinomial}
\end{align}
where $P(s',p'|s,p)$ denotes the conditional probability that, given a selected node of class $(s,p)$, a randomly selected neighbor belongs to class $(s',p')$. These conditional probabilities are calculated as the ratio between the number of links connecting nodes of classes $(s,p)$ and $(s',p')$, and the total number of links attached to nodes of class $(s,p)$.
Specific expressions of these conditional probabilities, as well as the dynamical closed equations for the set $\{\Delta, \Sigma, \rhop, \rhom, \rhos, \rhou\}$ that result from evaluating Eq.~\eqref{eq:HPA_general} together with Eqs.~\eqref{eq:p_flip},\eqref{eq:P_spnk}, \eqref{eq:multinomial}, are given in Eqs.\eqref{eq:PA_system} of \ref{app:PA}. They are the basis of our subsequent analysis.

\begin{figure}[ht!]
 \centering
\includegraphics[width=0.9\linewidth]{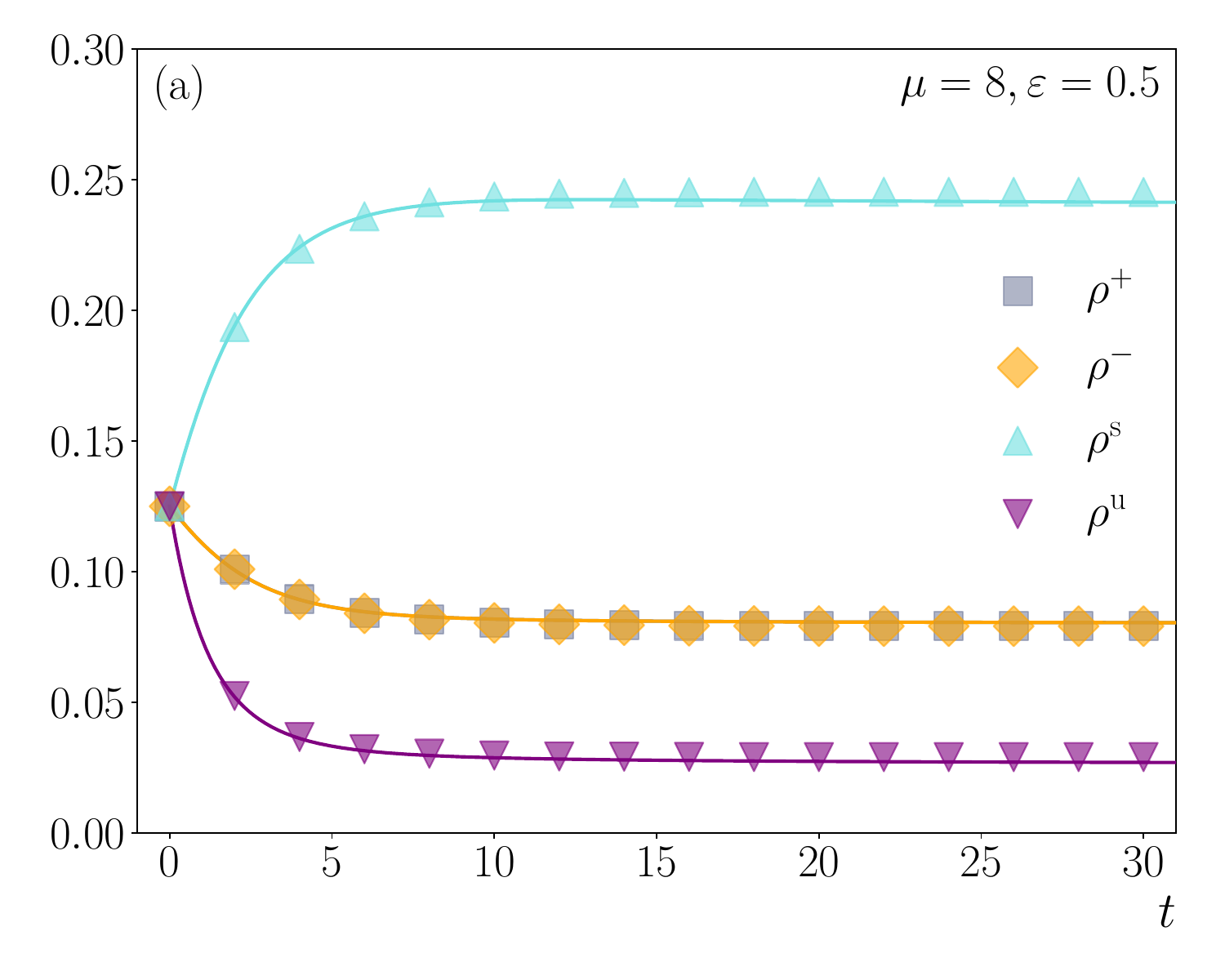} \includegraphics[width=0.9\linewidth]{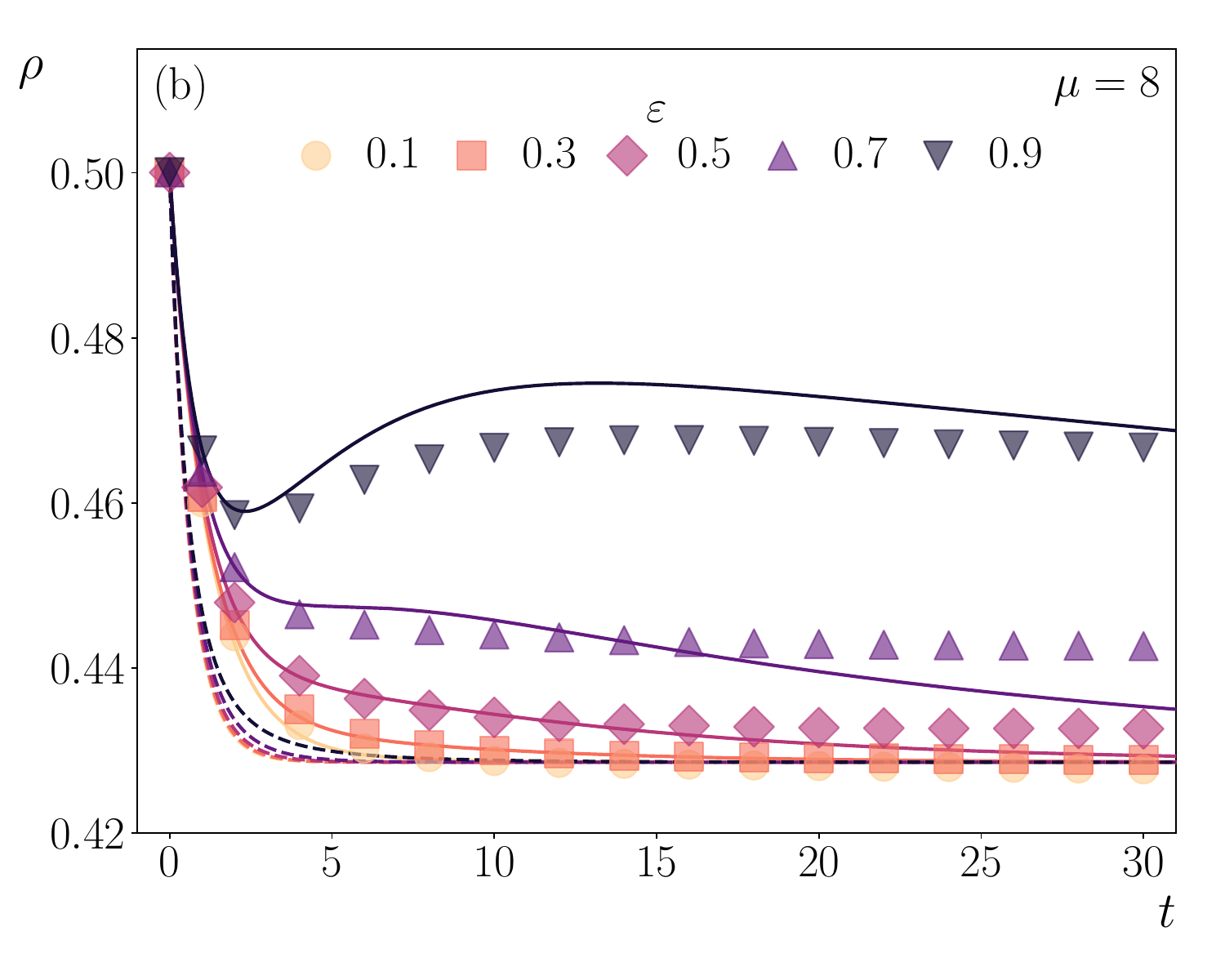}
 \caption{\textbf{Partisan voter model on uncorrelated networks. Time evolution of the densities of active links.} (a) Time evolution of the densities of active links $\rhop$, $\rhom$, $\rhos$, and $\rhou$, corresponding respectively to links between nodes with preference $p=+1$, preference $p=-1$, satisfied nodes, and unsatisfied nodes for $\varepsilon=0.5$. Data for $\rhop$ and $\rhom$ overlap. (b) Time evolution of the total density of active links $\rho$ for several values of $\varepsilon$, as indicated in the legend. Data for $\varepsilon=0.1$ and $\varepsilon=0.3$ nearly overlap. Symbols correspond to numerical simulations of the agent-based model on Erd\H{o}s--R\'enyi networks of size $N=10000$ and average degree $\mu=8$ while solid lines represent the numerical integration of the dynamical equations of the pair approximation Eqs.~\eqref{eq:PA_system}. In (b), dashed lines correspond to the numerical integration of the dynamical equations of the reduced pair approximation, Eqs.~\eqref{eq:PA_reduced_system}, where the dynamics is formulated solely in terms of the total density of active links $\rho$, see~\ref{app:PA_reduced}.}
 \label{fig:PA_t}
\end{figure}

The resulting dynamical system presents two non-hyperbolic saddle fixed points
\begin{equation} \label{eq:PA_solution_abs}
 (\Delta_\st, \Sigma_\st, \rhop_\st,\rhom_\st,\rhos_\st,\rhou_\st)=\left(\pm \frac 12, \frac{1}{2}, 0,0,0,0\right),
\end{equation}
corresponding to the two absorbing consensus states, analogous to the mean-field solution. 
Additionally, there exists a stable solution that corresponds to a coexistence of both states,
\begin{equation} \label{eq:PA_solution}
\begin{aligned}
&(\Delta_\st, \Sigma_\st, \rhop_\st,\rhom_\st,\rhos_\st,\rhou_\st)=\\
&\left(0, \frac{1+\varepsilon}{2},\frac{\xi}{4} (1-\varepsilon^2), \frac{\xi}{4} (1-\varepsilon^2), \frac{\xi}{4} (1+\varepsilon)^2, \frac{\xi}{4} (1-\varepsilon)^2\right),
\end{aligned}
\end{equation}
with
\begin{equation} \label{eq:xi}
\xi=\frac{\mu-2}{2(\mu-1)},
\end{equation}
where $\mu=\sum_k kP_k$ is the average degree of the network. Finite size fluctuations, not taken into account in this deterministic description, will eventually take the system to one of the absorbing states. Note that in the stationary state $\rhop_\st=\rhom_\st$ due to the symmetry of the model. By increasing $\varepsilon$, the density of active links between satisfied nodes $\rhos_\st$ is enhanced while the remaining contributions $\rhop_\st$, $\rhom_\st$, and $\rhou_\st$ are reduced. Moreover, the stationary values $\Delta_\st$ and $\Sigma_\st$ coincide with the mean-field prediction.

Importantly, the stationary total density of active links $\rho_\st$ is independent of the strength of the preference $\varepsilon$,
\begin{equation} \label{eq:PA_solution_rho}
\rho_\st=\rhop_\st+\rhom_\st+\rhos_\st+\rhou_\st=\xi.
\end{equation}
The factor $\xi$ coincides with that obtained in the pair approximation of the standard voter model~\cite{Vazquez_2008}. However, the interpretation in this case is different. In the standard voter model, the stationary states form a continuous family satisfying $\rho=\xi(1-m^2)$, parametrized by the conserved magnetization $m$. In contrast, in the partisan voter model the preference bias breaks the conservation of the magnetization, so that this continuous family collapses to a unique stable coexistence solution with $\rho_\st=\xi$. Therefore, the partisan bias does not modify the total density of active links in the stationary state, but rather redistributes it among the different types of active links.

In Fig.~\ref{fig:PA_ER} we compare the analytical predictions of the pair approximation with numerical simulations of the model defined in Sec.~\ref{sec:model} on Erd\H{o}s--R\'enyi networks. We show the stationary densities of active links as a function of the parameters $\mu$ and $\varepsilon$. Overall, we observe an excellent agreement between simulations and theory. 
For a fixed value of $\varepsilon=0.2$, Fig.~\ref{fig:PA_ER}(a) shows the dependence of the stationary densities $\rhop_\st,\rhom_\st,\rhos_\st,\rhou_\st$ on the average degree $\mu$. As $\mu$ increases, all densities converge towards their mean-field values.
For a fixed value of $\mu=8$, Fig.~\ref{fig:PA_ER}(b) shows the dependence on the strength of the preference $\varepsilon$. As $\varepsilon$ increases, alignment with individual preferences is favored, leading to a redistribution of the different types of active links. In particular, the density of active links connecting satisfied nodes increases, while the remaining contributions decrease.
Finally, Fig.~\ref{fig:PA_ER}(c) shows the total stationary density of active links $\rho_\st$ as a function of $\mu$ for different values of $\varepsilon$. In agreement with Eq.~\eqref{eq:PA_solution_rho}, the data displays an overall collapse for most values of $\varepsilon$, confirming that the total stationary activity is essentially independent of the partisan bias. Noticeable deviations appear only for values of $\varepsilon$ close to $1$. As Fig.~\ref{fig:PA_ER}(b) shows, in this regime the dominant contribution to $\rho_\st$ comes from $\rhos_\st$, while the other components become very small. Therefore, the residual discrepancy observed in Fig.~\ref{fig:PA_ER}(c) is mainly associated with the behavior of $\rhos_\st$ near the limit of strong bias.

It is worth noting that the same stationary solutions for $\Delta_\st,\Sigma_\st$ and $\rho_\st$ are also obtained within the reduced pair approximation discussed in~\ref{app:PA_reduced}, where the dynamics is formulated in terms of the total density of active links $\rho$ rather than the four partial densities $\rhop$, $\rhom$, $\rhos$, and $\rhou$.
However, the detailed pair approximation has two important advantages over the reduced scheme of ~\ref{app:PA_reduced}: it explicitly resolves the redistribution of activity among the different classes of active links, and it provides a substantially more accurate description of the transient dynamics. This is illustrated in Fig.~\ref{fig:PA_t}, where we compare the time evolution of the densities of active links obtained from numerical simulations of the agent-based model on Erd\H{o}s--R\'enyi networks with the predictions of both pair approximation schemes. Fig.~\ref{fig:PA_t}(a) shows, for $\mu=8$ and $\varepsilon=0.5$, the temporal evolution of the four densities $\rhop(t)$, $\rhom(t)$, $\rhos(t)$, and $\rhou(t)$. We find excellent agreement between simulations and the detailed pair approximation solution. Starting from a homogeneous initial condition, the dynamics quickly converges to the the stationary state. In particular, $\rhos(t)$ increases and becomes the dominant contribution, whereas $\rhop(t)$, $\rhom(t)$, and $\rhou(t)$ relax towards smaller stationary values. 

Fig.~\ref{fig:PA_t}(b) compares the time evolution of the total density of active links $\rho(t)$ for several values of $\varepsilon$ at fixed $\mu=8$. The detailed pair approximation (solid lines) provides a clearly more precise description of the transient dynamics than the reduced pair approximation (dashed lines). In particular, for large values of the preference $\varepsilon$ the reduced approximation fails to capture the non-monotonic relaxation observed in the simulations, whereas the detailed scheme reproduces it much more accurately. This shows that resolving the different classes of active links is not only necessary to describe how they are evolving in time, but also crucial for an precise quantitative account of the relaxation towards the steady state.

A natural next step is to compare the predictions of the pair approximation with numerical results on Barab\'asi--Albert networks. In the next section, we consider this case together with a broader analysis of networks generated by preference-based attachment.

\begin{figure}[t]
 \centering 
\includegraphics[width=0.9\linewidth]{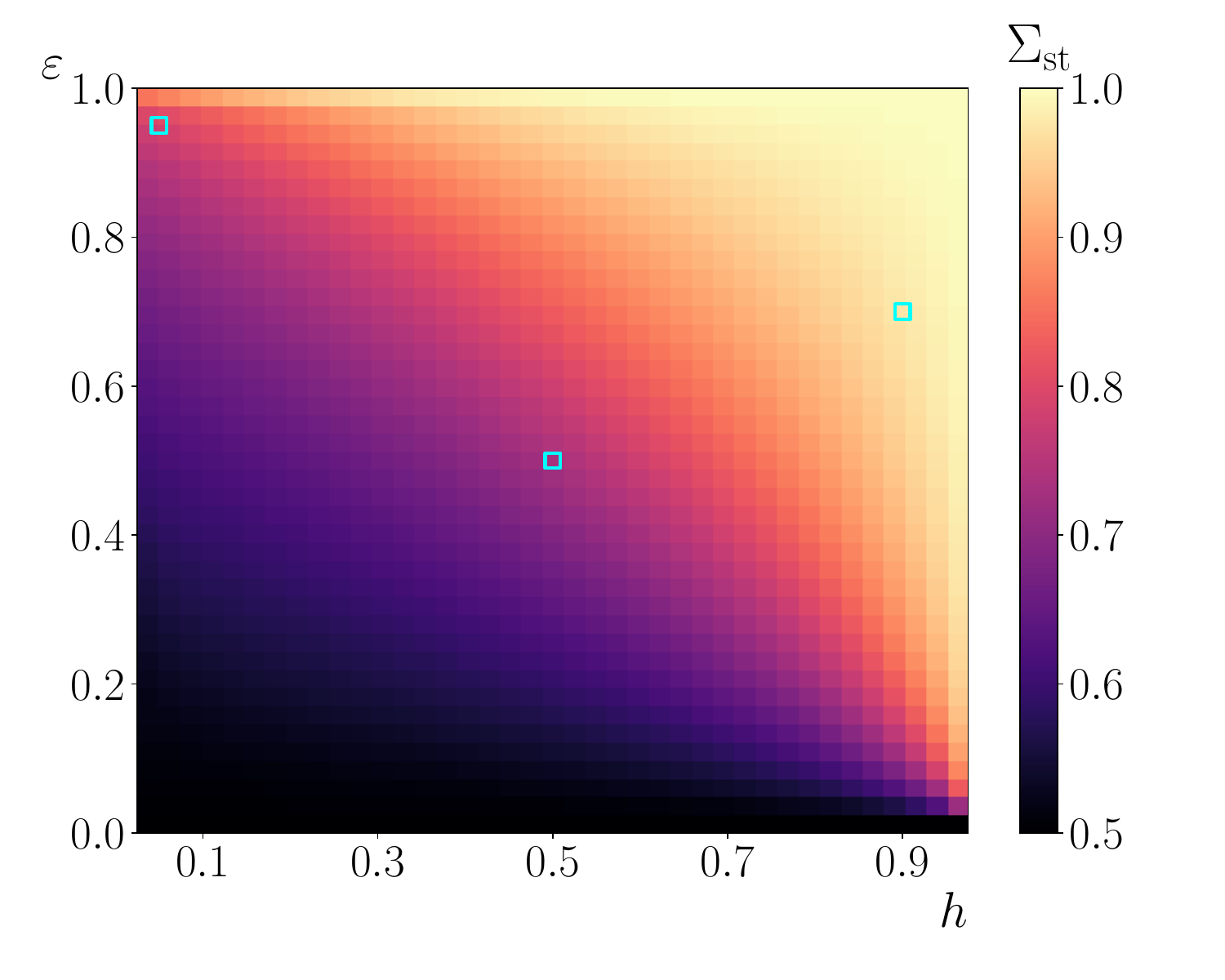}
 \caption{\textbf{Partisan voter model on networks with preference-based attachment. Density of satisfied agents.} Heat map of the stationary values of the density of satisfied agents $\Sigma_\st$ in the parameter space $(h, \varepsilon)$. Results correspond to numerical simulations of the agent-based model on networks of size $N=10000$ constructed with the Barab\'asi--Albert-homophily model with $m_0=4$ and $m=4$, resulting in a network of average degree $\mu=8$. Cyan squares indicate the parameter values $(h,\varepsilon)$ corresponding to the representative network snapshots shown in Fig.~\ref{fig:BAh_rho} (c)--(e).}
 \label{fig:BAh_sigma}
\end{figure}

\begin{figure*}[ht!]
 \centering
\includegraphics[width=0.95\linewidth]{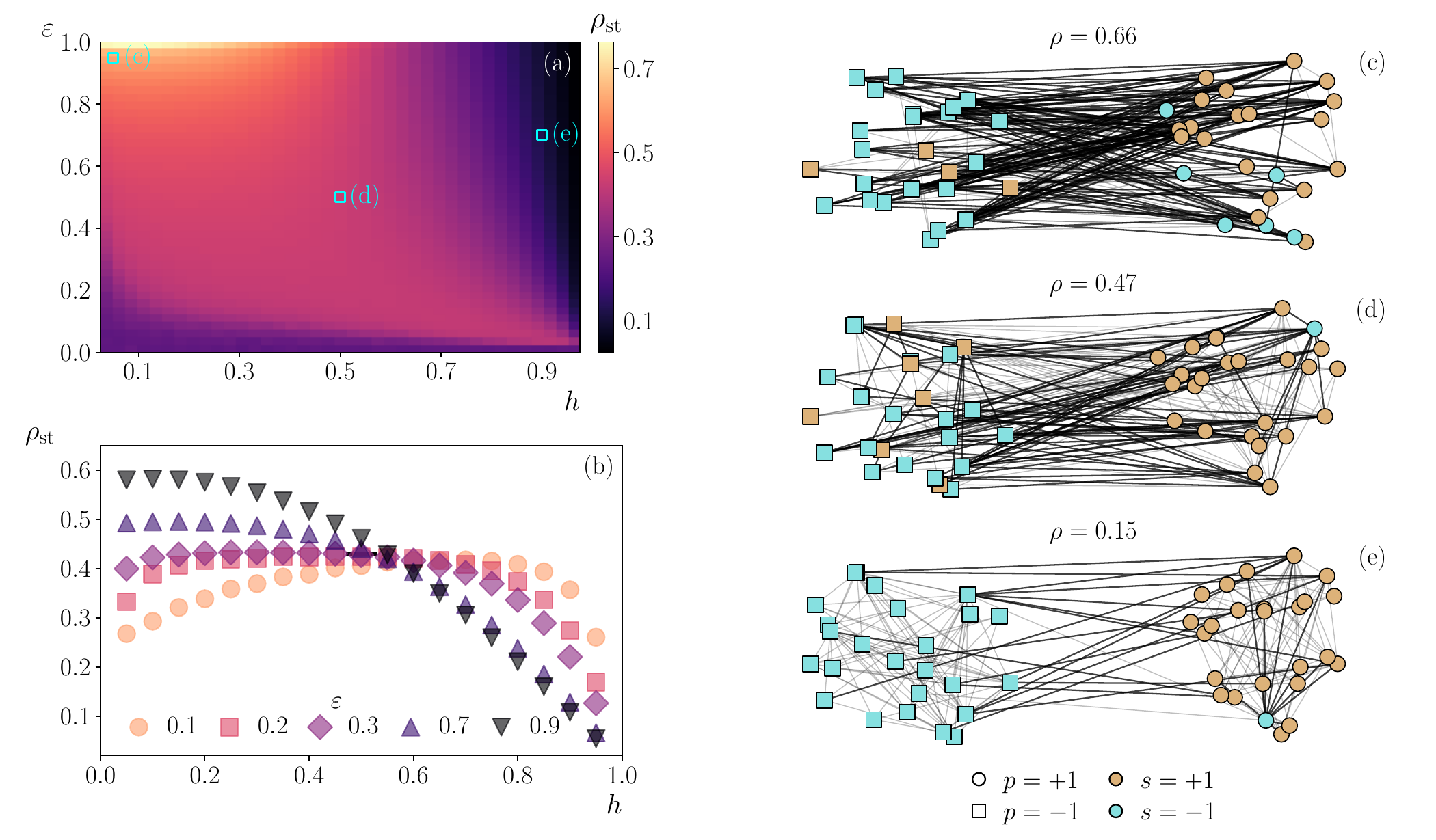} 
 \caption{\textbf{Partisan voter model on networks with preference-based attachment. Total density of active links.} (a) Heat map of the stationary values of the total density of active links $\rho_\st$ in the parameter space $(h,\varepsilon)$. (b) Stationary values of the total density of active links $\rho_\st$ as a function of the homophily parameter $h$ for several values of $\varepsilon$, as indicated in the legend. The curves correspond to horizontal cuts of the heat map shown in panel~(a). The short horizontal line at $h=0.5$ indicates the pair-approximation prediction, Eq.~\eqref{eq:PA_solution}. Results in panels~(a,b) are obtained from numerical simulations of the agent-based model on networks of size $N=10000$ generated with the Barab\'asi--Albert homophily model with $m_0=4$ and $m=4$, yielding an average degree $\mu=8$. (c)--(e) Representative stationary snapshots of networks of size $N=50$ with average degree $\mu=8$ for the parameter values $(h,\varepsilon)$ marked by cyan squares in panel~(a).}
 \label{fig:BAh_rho}
\end{figure*}

\section{Partisan voter model on networks with preference-based attachment} \label{sec:homophily}
In many social systems, interaction patterns are not random, but are conditioned by homophily, i.e., the tendency of individuals to connect preferentially with others sharing similar attributes~\cite{McPherson_2001,Karimi_2018,Karimi_2019,diaz_homophily}. In this section, we move beyond uncorrelated random networks and study the partisan voter model on networks with preference-dependent connectivity. Specifically, we consider networks generated with the Barab\'asi--Albert homophily model~\cite{Karimi_2018,Karimi_2019}, a generalization of the Barab\'asi--Albert model~\cite{BA}. Simple and complex contagion processes have been studied in these kinds of networks~\cite{diaz_homophily}. In this paper, we will assume that the population is divided into two groups of equal size, each consisting of agents with the same preference.

The network is constructed as follows. We start from an initial fully connected seed of $m_0=4$ nodes, two with preference $p=+1$ and two with preference $p=-1$. The remaining, initially disconnected, $N-4$ nodes are assigned preferences at random, with half assigned $p=+1$ and the other half $p=-1$. At each iteration step, a node $j$ is selected at random from those not yet included in the network and is connected to $m=4$ pre--existing nodes such that each target node $i$ is selected with probability
\begin{equation}
 P_{j\to i} = \frac{h_{ij} k_i}{\sum_{\ell} h_{\ell j} k_\ell},
\end{equation}
where $\ell$ is the total set of pre-existing nodes, $k_i$ is the degree of node $i$, and $h_{ij}$ introduces preference-based attachment according to
\begin{equation}
 h_{ij} =
 \begin{cases}
 h, & \text{if } p_i = p_j, \\
 1 - h, & \text{if } p_i \neq p_j.
 \end{cases}
\end{equation}
The resulting network presents a degree distribution $P(k)=k^{-\gamma}$, with $\gamma=3$, and average degree $\mu=2m=8$. The homophily parameter $h \in [0,1]$ controls the tendency to connect to nodes with the same preference. For $h < 1/2$, nodes preferentially connect to nodes with opposite preference, resulting in \textit{heterophilic} networks. For $h > 1/2$, connections are biased towards nodes with the same preference, leading to \textit{homophilic} networks. The case $h=1/2$ recovers the standard Barab\'asi--Albert network~\cite{BA}. In the limiting case $h=1$, links are only formed between nodes with identical preference, and the network splits into two almost disconnected components linked by the initial set of $m_0$ nodes, one for each preference group. 

Since the structural correlations induced by homophily are not captured by the pair approximation, we characterize the stationary state of the system numerically. To this end, we perform computer simulations of the partisan voter model on Barab\'asi--Albert homophily networks for several values of the homophily parameter $h$ and the preference strength $\varepsilon$.

% Discussion about the global state of the system
When analyzing the global steady state of the system, we first note that, for finite networks, the dynamics fluctuates around a coexistence state with $\Delta=0$ and no macroscopic dominance of either opinion, until a finite-size fluctuation eventually drives the system to one of the two absorbing states. However, this condition alone does not determine how coexistence is organized with respect to the agents' intrinsic preferences. This information is captured by the stationary density of agents that are in their preferred state $\Sigma_\st$.
% Sigma
In Fig.~\ref{fig:BAh_sigma}, we plot the heat map of $\Sigma_\st$ in the parameter space $(h,\varepsilon)$. As expected, increasing $\varepsilon$ systematically increases the fraction of agents aligned with their preferred opinion. The effect of the network structure is weaker but still significant. Heterophilic networks ($h<1/2$) generate frustration between social influence and intrinsic bias, reducing the overall level of satisfaction. On the other hand, homophilic networks ($h>1/2$) favor larger values of $\Sigma_\st$, since agents are predominantly surrounded by neighbors sharing the same preference. In the strongly homophilic limit $h\to1$, then $\Sigma_\st\to1$, consistently with the emergence of segregated communities internally ordered according to their preferred state.

% Total rho
We now focus on how coexistence is organized across the network. The stationary behavior of the total density of active links $\rho_\st$ is shown in Fig.~\ref{fig:BAh_rho}. Figure~\ref{fig:BAh_rho}(a) displays the heat map of $\rho_\st$ in the parameter space $(h,\varepsilon)$, Fig.~\ref{fig:BAh_rho}(b) shows horizontal cuts of the heat map for several values of $\varepsilon$, and Fig.~\ref{fig:BAh_rho}(c)--(e) displays representative snapshots of the stationary state of the network.

The heat map reveals three different qualitative stationary regimes. The first one corresponds to the heterophilic regime ($h<1/2$) and large values of $\varepsilon$, where the system reaches highly disordered stationary configurations with very large densities of active links. This regime originates from the competition between heterophilic connectivity, which favors interactions across opposite preferences, and the strong tendency of agents to remain aligned with their intrinsic preference. A representative configuration in this regime is illustrated in Fig.~\ref{fig:BAh_rho}(c). We observe the presence of many neighboring nodes that remain in opposite states despite the segregation in preferences.

The second regime corresponds to a broad intermediate region of the parameter space characterized by values $\rho_\st\sim0.5$. It corresponds to coexistence between the two states with a moderate degree of local ordering, namely clusters of intermediate size. A representative configuration in this regime is shown in Fig.~\ref{fig:BAh_rho}(d).

The third regime is characterized by values $\rho_\st<0.3$, indicating coexistence with strong local ordering and large domains of each state. This regime is observed either for sufficiently small $\varepsilon$ at fixed $h$, or for sufficiently large $h$ at fixed $\varepsilon$. In the strongly homophilic limit $h\to1$, it is $\rho_\st\to0$, reflecting the fragmentation of the network into almost disconnected communities sharing the same preference, within which local consensus is easily achieved. This organization is illustrated in Fig.~\ref{fig:BAh_rho}(e), where the network becomes effectively segregated into communities internally ordered according to their preferred state.

The snapshots of Fig.~\ref{fig:BAh_rho}(c),(d),(e) corresponding to the three aforementioned regimes are also consistent with the behavior of $\Sigma_\st$: heterophilic configurations display a stronger mismatch between preferences and states, whereas homophilic networks exhibit communities in which both variables are largely aligned.

% Cuts
The horizontal cuts shown in Fig.~\ref{fig:BAh_rho}(b) further clarify the dependence of $\rho_\st$ on the homophily parameter $h$. The curves display a crossover controlled by the preference strength $\varepsilon$: for small values of $\varepsilon$, $\rho_\st$ depends nonmonotonically on $h$, reaching a maximum at intermediate homophily, whereas for larger values of $\varepsilon$, $\rho_\st$ decreases monotonically with $h$. Thus, as $\varepsilon$ increases, the system crosses over from a regime in which intermediate homophily enhances activity to one in which homophily suppresses it throughout the whole range. Consistently, increasing $\varepsilon$ has opposite effects depending on the structural regime: it enhances activity in heterophilic networks, while it suppresses it in homophilic networks.

\begin{figure}[ht!]
 \centering
\includegraphics[width=0.9\linewidth]{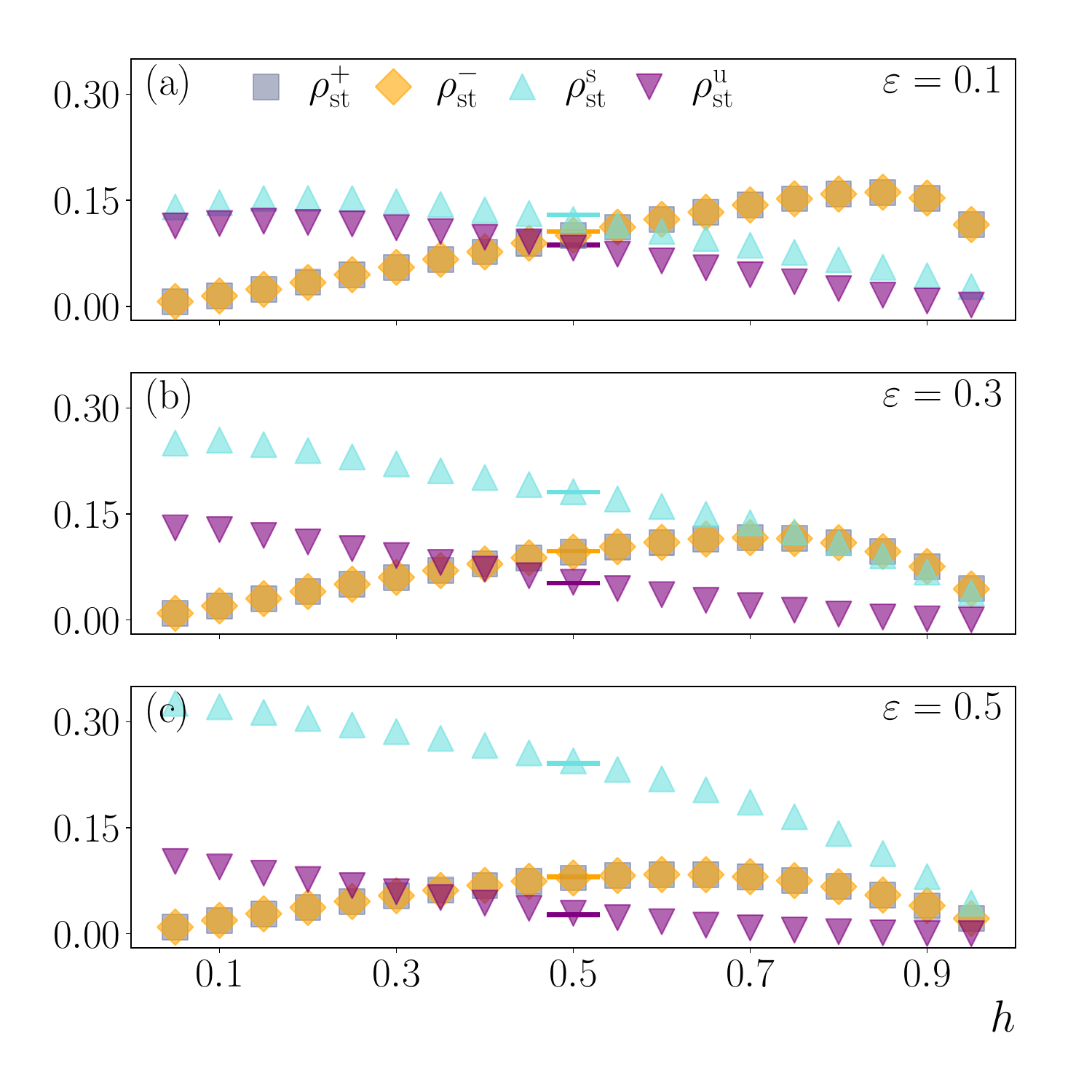} 
 \caption{\textbf{Partisan voter model on networks with preference-based attachment. Stationary state of the densities of active links.} Stationary values of the densities of active links $\rhop_\st$, $\rhom_\st$, $\rhos_\st$, and $\rhou_\st$, corresponding respectively to links between nodes with preference $p=+1$, preference $p=-1$, satisfied nodes, and unsatisfied nodes, as a function of the homophily parameter $h$, for $\varepsilon=0.1, 0.3, 0.5$. Data for $\rhop_\st$ and $\rhom_\st$ overlap, reflecting the symmetry between the two preference groups. Symbols correspond to numerical simulations of the agent-based model on networks of size $N=10000$ constructed with the Barab\'asi--Albert-homophily model with $m_0=4$ and $m=4$, resulting in a network of average degree $\mu=8$. Short horizontal lines located at $h=0.5$ correspond to the prediction of the pair approximation, Eq.~\eqref{eq:PA_solution}.}
 \label{fig:rhos_BAh}
\end{figure}

%Discussion about the rhos
To understand the microscopic origin of the global behavior described above, we now analyze how the total activity is distributed among the different classes of active links. Figure~\ref{fig:rhos_BAh} shows the stationary values of the different densities of active links as a function of the homophily parameter $h$ for increasing values of the preference strength $\varepsilon$. Due to the symmetry between the two preference groups, the densities $\rhop_\st$ and $\rhom_\st$ coincide for all values of $h$. For Barab\'asi--Albert networks, corresponding to $h=1/2$, we observe good overall agreement with the predictions of the pair approximation, Eq.~\eqref{eq:PA_solution}.
The four classes of active links respond qualitatively differently to increasing $h$. The densities $\rhop_\st$ and $\rhom_\st$ display a nonmonotonic dependence on $h$, reaching a maximum at intermediate values within the homophilic regime, $h>1/2$. By contrast, $\rhos_\st$ and $\rhou_\st$ generally decrease with $h$, except for small values of $\varepsilon$, where they exhibit a shallow maximum at low values of $h$.
Increasing $\varepsilon$ strongly redistributes the relative weight of the different contributions. In particular, $\rhos_\st$ becomes the dominant contribution over the whole range of $h$, whereas $\rhou_\st$ remains relevant only in the heterophilic regime $h<1/2$ and rapidly decreases as $h$ grows. At the same time, the maximum of $\rho^{\pm}_\st$ shifts towards smaller values of $h$.

%Closing section
Altogether, these results show that in networks with preference-based attachment, the composition of active links is strongly reorganized and, depending on the value of $\varepsilon$, the overall stationary level of activity can be either enhanced or suppressed. This behavior reveals a nontrivial competition between the preference bias in the dynamics, a global mechanism, and the local effect of the structural organization induced by homophily. 

\section{Discussion} \label{sec:conclusions}
In this work, we have studied the partisan voter model on complex networks by combining a pair approximation analytical treatment with numerical simulations. Since each agent has a fixed preference for one of the two states, not all active links are dynamically equivalent. Consequently, the total density of active links $\rho$ is insufficient to characterize the dynamics. We therefore decompose $\rho$ into four contributions, $\rho^{+}$, $\rho^{-}$, $\rho^{s}$, and $\rho^{u}$, corresponding respectively to links between nodes with preference $p=+1$, preference $p=-1$, satisfied nodes, and unsatisfied nodes. This decomposition is the key ingredient of the theory developed in our work.

For uncorrelated random networks, we developed a pair approximation in terms of the variables $\{\Delta, \Sigma, \rhop, \rhom, \rhos, \rhou\}$, which resolves the different classes of active links and yields a closed description of the dynamics. This theory predicts a stable stationary coexistence state with $\Delta_\st=0$ and $\Sigma_\st=\tfrac 12(1+\varepsilon)$, in agreement with the mean-field theory, and a dependence of the densities of active links on the preference strength $\varepsilon$. In particular, as $\varepsilon$ increases, stationary density active links between satisfied nodes, $\rhos_\st$, become more prevalent, whereas the remaining contributions decrease. The excellent agreement between the pair approximation and numerical simulations of the agent-based model on Erd\H{o}s--R\'enyi networks shows that the theory captures the stationary state accurately over a broad parameter range, with appreciable deviations only as $\varepsilon\to 1$.
Interestingly, the total stationary density of active links is independent of $\varepsilon$, $\rho_\st=(\mu-2)/[2(\mu-1)]$. While the same stationary solution is also obtained within the reduced pair approximation formulated solely in terms of the total density of active links $\rho_\st$, resolving the partial densities is essential to accurately capture the transient dynamics. Thus, in uncorrelated networks, preferences do not change the total stationary density of active links, but rather reorganizes them among their different classes.

%Discussion BAh
The situation changes qualitatively when the network structure is correlated with the preferences. In Barab\'asi--Albert homophily networks, the same quenched preferences that bias the dynamics also shape the connectivity pattern. As a result, the stationary coexistence state becomes strongly dependent on the interplay between the preference strength $\varepsilon$ and the homophily parameter $h$. For weak partisan bias, the total density of active links displays a nonmonotonic dependence on homophily, with a maximum at intermediate values of $h$. For stronger bias, by contrast, the stationary activity decreases monotonically with $h$. The system therefore crosses over from highly disordered heterophilic configurations with large densities of active links to segregated homophilic communities with strong local ordering and almost no active links. Equivalently, increasing $\varepsilon$ has opposite effects depending on the structural regime: it enhances $\rho_\st$ in heterophilic networks but suppresses it in homophilic ones. Thus, the effect of partisanship cannot be understood independently of the network structure.

This global phenomenology originates from a nontrivial reorganization of the different classes of active links induced by homophily. The four contributions respond differently to the homophily parameter $h$: $\rhou_\st$ is progressively suppressed as the network becomes more homophilic, whereas $\rhop_\st$ and $\rhom_\st$ display a nonmonotonic dependence on $h$, reaching a maximum at intermediate values. At the same time, increasing $\varepsilon$ enhances the relative importance of $\rhos_\st$, which becomes the dominant contribution over most of the parameter range. Homophily therefore affects not only the composition of active links, but also their total stationary density, by changing the relative abundance of contacts within and across preference groups. For $h=1/2$, corresponding to standard Barab\'asi--Albert networks, we observe good agreement between numerical simulations and the pair approximation predictions.

Overall, our results identify two distinct roles of partisan preferences. The parameter $\varepsilon$ acts at the dynamical level: it biases the microscopic transition probabilities, breaks the conservation of magnetization, and determines how disagreement is distributed among the different classes of active links. By contrast, the homophily parameter $h$ acts at the structural level: it introduces local preference correlations in the network by biasing link formation towards agents with the same or opposite preference. The interplay between these two mechanisms explains why the same partisan bias can produce qualitatively different collective behaviors depending on the underlying network structure. This highlights the importance of distinguishing dynamical biases from structural correlations when modeling opinion dynamics in social systems.

\textbf{Acknowledgments:}
Partial financial support has been received from Grant PID2024-157493NB-C21 funded by MICIU/AEI/10.13039/501100011033 and by ERDF, EU, and the Mar{\'\i}a de Maeztu Program for units of Excellence in R\&D, grant CEX2021-001164-M.

\bibliographystyle{unsrt}
\bibliography{main}

\begin{thebibliography}{10}

\bibitem{Castellano_2009}
Claudio Castellano, Santo Fortunato, and Vittorio Loreto.
\newblock Statistical physics of social dynamics.
\newblock {\em Rev. Mod. Phys.}, 81:591, 2009.

\bibitem{Starnini_2025}
Michele Starnini, Fabian Baumann, Tobias Galla, David Garcia, Gerardo Iñiguez, Márton Karsai, Jan Lorenz, and Katarzyna Sznajd-Weron.
\newblock Opinion dynamics: Statistical physics and beyond.
\newblock {\em arXiv preprint arXiv:2507.11521}, 2025.

\bibitem{Clifford_1973}
Peter Clifford and Aidan Sudbury.
\newblock A model for spatial conflict.
\newblock {\em Biometrika}, 60(3):581--588, 1973.

\bibitem{Holley_1975}
Richard~A. Holley and Thomas~M. Liggett.
\newblock Ergodic theorems for weakly interacting infinite systems and the voter model.
\newblock {\em The Annals of Probability}, 3(4):643--663, 1975.

\bibitem{Sood_2005}
Vishal Sood and Sidney Redner.
\newblock Voter model on heterogeneous graphs.
\newblock {\em Phys. Rev. Lett.}, 94:178701, 2005.

\bibitem{Suchecki_2005}
Krzysztof Suchecki, V{\'i}ctor~M. Egu{\'i}luz, and Maxi San~Miguel.
\newblock Voter model dynamics in complex networks: Role of dimensionality, disorder, and degree distribution.
\newblock {\em Phys. Rev. E}, 72:036132, 2005.

\bibitem{Sood_2008}
Vishal Sood, Tibor Antal, and Sidney Redner.
\newblock Voter models on heterogeneous networks.
\newblock {\em Phys. Rev. E}, 77:041121, 2008.

\bibitem{Vazquez_2008}
Federico Vazquez and V{\'\i}ctor~M Egu{\'\i}luz.
\newblock Analytical solution of the voter model on uncorrelated networks.
\newblock {\em New Journal of Physics}, 10:063011, 2008.

\bibitem{FernandezGracia_2014}
Juan Fern{\'a}ndez-Gracia, Krzysztof Suchecki, Jos{\'e}~J. Ramasco, Maxi San~Miguel, and V{\'i}ctor~M. Egu{\'i}luz.
\newblock Is the voter model a model for voters?
\newblock {\em Phys. Rev. Lett.}, 112:158701, 2014.

\bibitem{Hammal_2005}
Omar Al~Hammal, Hugues Chat\'e, Ivan Dornic, and Miguel~A. Mu\~noz.
\newblock Langevin description of critical phenomena with two symmetric absorbing states.
\newblock {\em Phys. Rev. Lett.}, 94:230601, 2005.

\bibitem{Llabres_2026}
Jaume Llabr\'es, Maxi San~Miguel, and Ra\'ul Toral.
\newblock Universality of noise-induced transitions in nonlinear voter models.
\newblock {\em Phys. Rev. Res.}, 8:013015, 2026.

\bibitem{Mobilia_2003}
Mauro Mobilia.
\newblock Does a single zealot affect an infinite group of voters?
\newblock {\em Phys. Rev. Lett.}, 91:028701, 2003.

\bibitem{Masuda_2010}
Naoki Masuda, Nicolas Gibert, and Sidney Redner.
\newblock Heterogeneous voter models.
\newblock {\em Phys. Rev. E}, 82:010103, 2010.

\bibitem{Vazquez_2010}
Federico Vazquez, Xavier Castelló, and Maxi San~Miguel.
\newblock Agent based models of language competition: macroscopic descriptions and order–disorder transitions.
\newblock {\em Journal of Statistical Mechanics: Theory and Experiment}, 2010:P04007, 2010.

\bibitem{Masuda_2011}
Naoki Masuda and Sidney Redner.
\newblock Can partisan voting lead to truth?
\newblock {\em Journal of Statistical Mechanics: Theory and Experiment}, 2011:L02002, 2011.

\bibitem{Redner_2019}
Sidney Redner.
\newblock Reality-inspired voter models: A mini-review.
\newblock {\em Comptes Rendus. Physique}, 20(4):275--292, 2019.

\bibitem{Czaplicka_2022}
Agnieszka Czaplicka, Christos Charalambous, Raul Toral, and Maxi {San Miguel}.
\newblock Biased-voter model: How persuasive a small group can be?
\newblock {\em Chaos, Solitons \& Fractals}, 161:112363, 2022.

\bibitem{Llabres_2023}
Jaume Llabr\'es, Maxi San~Miguel, and Ra\'ul Toral.
\newblock Partisan voter model: Stochastic description and noise-induced transitions.
\newblock {\em Phys. Rev. E}, 108:054106, 2023.

\bibitem{Suchecki_Euro_2005}
Krzysztof Suchecki, V{\'i}ctor~M. Egu{\'i}luz, and Maxi San~Miguel.
\newblock Conservation laws for the voter model in complex networks.
\newblock {\em Europhysics Letters}, 69:228, 2004.

\bibitem{Castellano_2005}
Claudio Castellano, Vittorio Loreto, Alain Barrat, Federico Cecconi, and Domenico Parisi.
\newblock Comparison of voter and glauber ordering dynamics on networks.
\newblock {\em Phys. Rev. E}, 71:066107, 2005.

\bibitem{Gleeson_2011}
James~P. Gleeson.
\newblock High-accuracy approximation of binary-state dynamics on networks.
\newblock {\em Phys. Rev. Lett.}, 107:068701, 2011.

\bibitem{Kuehn_2016}
Christian Kuehn.
\newblock {Moment closure—a brief review}.
\newblock In Eckehard Schöll, Sabine~H.L. Klapp, and Philipp Hövel, editors, {\em Control of Self-Organizing Nonlinear Systems}, Understanding Complex Systems, pages 253--271. Springer, 2016.

\bibitem{Peralta_2018}
Antonio~F. Peralta, Adriàn Carro, Maxi {San Miguel}, and Raúl Toral.
\newblock Analytical and numerical study of the non-linear noisy voter model on complex networks.
\newblock {\em Chaos: An Interdisciplinary Journal of Nonlinear Science}, 28:075516, 2018.

\bibitem{Peralta_Sto_2018}
Antonio~F. Peralta, Adrián Carro, Maxi {San Miguel}, and Raúl Toral.
\newblock Stochastic pair approximation treatment of the noisy voter model.
\newblock {\em New Journal of Physics}, 20:103045, 2018.

\bibitem{Ramirez_2024}
Luc\'{\i}a~S. Ramirez, Federico Vazquez, Maxi San~Miguel, and Tobias Galla.
\newblock Ordering dynamics of nonlinear voter models.
\newblock {\em Phys. Rev. E}, 109:034307, 2024.

\bibitem{McPherson_2001}
Miller McPherson, Lynn Smith-Lovin, and James~M. Cook.
\newblock Birds of a feather: Homophily in social networks.
\newblock {\em Annual Review of Sociology}, 27:415, 2001.

\bibitem{Flache_2011}
Andreas Flache and Michael~W. Macy.
\newblock Small worlds and cultural polarization.
\newblock {\em The Journal of Mathematical Sociology}, 35:146, 2011.

\bibitem{Mas_2013}
Michael M{\"a}s and Andreas Flache.
\newblock Differentiation without distancing. explaining bi-polarization of opinions without negative influence.
\newblock {\em PLOS ONE}, 8(11):e74516, 2013.

\bibitem{Karimi_2018}
Fariba Karimi, Mathieu G{\'e}nois, Claudia Wagner, Philipp Singer, and Markus Strohmaier.
\newblock Homophily influences ranking of minorities in social networks.
\newblock {\em Scientific Reports}, 8(1):11077, 2018.

\bibitem{Karimi_2019}
Eun Lee, Fariba Karimi, Claudia Wagner, Hang-Hyun Jo, Markus Strohmaier, and Mirta Galesic.
\newblock Homophily and minority-group size explain perception biases in social networks.
\newblock {\em Nature Human Behaviour}, 3(10):1078--1087, 2019.

\bibitem{diaz_homophily}
Fernando Diaz-Diaz, Maxi San~Miguel, and Sandro Meloni.
\newblock Echo chambers and information transmission biases in homophilic and heterophilic networks.
\newblock {\em Scientific Reports}, 12(1):9350, 2022.

\bibitem{BA}
Albert-László Barabási and Réka Albert.
\newblock Emergence of scaling in random networks.
\newblock {\em Science}, 286(5439):509--512, 1999.

\end{thebibliography}

\appendix
\widetext
\section{Pair approximation for the partisan voter model} \label{app:PA} 

To compute the evolution equations of the pair approximation developed in Sec.~\ref{sec:PA} we need to combine Eqs.~\eqref{eq:HPA_general}--\eqref{eq:multinomial} together with the conditional probabilities $P(s',p'|s,p)$ that, given a selected node of class $(s,p)$, a randomly selected neighbor belongs to class $(s',p')$. These probabilities are computed as the ratio between the number of links $\varphi\big((s,p),(s',p')\big)\mu N / 2$ connecting nodes of classes $(s,p)$ and $(s',p')$, and the total number of links $ x^p_s\mu N$ attached to nodes of class $(s,p)$
\begin{equation}
P(s',p'|s,p)=\frac{\varphi\big((s,p),(s',p')\big)}{2 x_s^p},
\label{eq:conditional_generic}
\end{equation}
where $\varphi\big((s,p),(s',p')\big)$ denotes the density of links connecting nodes of classes $(s,p)$ and $(s',p')$. Since links are undirected, this quantity is symmetric, namely
$\varphi\big((s,p),(s',p')\big)=\varphi\big((s',p'),(s,p)\big)$. These probabilities fulfill the normalization condition
\begin{equation}
\sum_{s'=\pm}\sum_{p'=\pm} P(s',p'|s,p)=1.
\end{equation}
By definition, it follows
\begin{subequations}
\begin{align}
\varphi\big((+,+),(-,+)\big)&=\rhop, \\
\varphi\big((+,-),(-,-)\big)&=\rhom, \\
\varphi\big((+,+),(-,-)\big)&=\rhos, \\
\varphi\big((+,-),(-,+)\big)&=\rhou,
\end{align}
\end{subequations}
which leads to the determination of $8$ out of the $16$ possible conditional probabilities
\begin{eqnarray}
P(-,+|+,+) &= \dfrac\rhop{2x_+^+}, \quad \quad P(-,-|+,+) = \dfrac\rhos{2x_+^+}, 
\quad \quad P(+,+|-,+) = \dfrac\rhop{2x_-^+}, \quad \quad P(+,-|-,+) = \dfrac\rhou{2x_-^+}, \nonumber\\
P(-,+|+,-) &= \dfrac\rhou{2x_+^-}, \quad \quad P(-,-|+,-) = \dfrac\rhom{2x_+^-}, \quad \quad P(+,-|-,-) = \dfrac\rhom{2x_-^-}, \quad \quad P(+,+|-,-) = \dfrac\rhos{2x_-^-}.
\end{eqnarray}
The remaining conditional probabilities corresponding to inactive links, i.e., neighbors with the same state $s'=s$, are approximated by distributing the remaining probability among the two preference classes $p'=\pm$ proportionally to their global densities within state $s$, namely
\begin{equation}
P(s,p'|s,p)=
\left(
1-\sum_{\tilde p=\pm}P(-s,\tilde p|s,p)
\right)
\frac{x_s^{p'}}{x_s^+ + x_s^-}.
\label{eq:inert_closure}
\end{equation}
This closure assumes that, conditioned on having the same state as the focal node, the preference of a neighbor is sampled according to its global abundance among nodes in state $s$, thereby neglecting higher-order correlations beyond pairs.

The final ingredient is the elementary variation $\delta X(s\to -s,\nb)$ in the observable $X$ when the flip occurs. This is summarized in Table~\ref{tab:elementary_changes} for each of the observables $\{\Delta, \Sigma, \rhop, \rhom, \rhos, \rhou\}$ under each possible flip $(s,p)\to(-s,p)$.

\begin{table}[h]
\centering
\begin{tabular}{c|c|c|c|c|c|c|c}
\hline
Transition & $\delta\Delta$ & $\delta\Sigma$ & $\delta\rhop$ & $\delta\rhom$ & $\delta\rhos$ & $\delta\rhou$ & $\delta\rho$ \\
\hline 
$(+,+)\to(-,+)$
& $-\dfrac{1}{N}$
& $-\dfrac{1}{N}$
& $\dfrac{n_+^+ - n_-^+}{E}$
& $0$
& $-\dfrac{n_-^-}{E}$
& $\dfrac{n_+^-}{E}$
& $\dfrac{n_+^+ + n_+^- - n_-^+ - n_-^-}{E}$
\\[3mm]

$(-,+)\to(+,+)$
& $\dfrac{1}{N}$
& $\dfrac{1}{N}$
& $\dfrac{-\,n_+^+ + n_-^+}{E}$
& $0$
& $\dfrac{n_-^-}{E}$
& $-\dfrac{n_+^-}{E}$
& $\dfrac{-\,n_+^+ - n_+^- + n_-^+ + n_-^-}{E}$
\\[3mm]

$(+,-)\to(-,-)$
& $-\dfrac{1}{N}$
& $\dfrac{1}{N}$
& $0$
& $\dfrac{n_+^- - n_-^-}{E}$
& $\dfrac{n_+^+}{E}$
& $-\dfrac{n_-^+}{E}$
& $\dfrac{n_+^+ + n_+^- - n_-^+ - n_-^-}{E}$
\\[3mm]

$(-,-)\to(+,-)$
& $\dfrac{1}{N}$
& $-\dfrac{1}{N}$
& $0$
& $\dfrac{-\,n_+^- + n_-^-}{E}$
& $-\dfrac{n_+^+}{E}$
& $\dfrac{n_-^+}{E}$
& $\dfrac{-\,n_+^+ - n_+^- + n_-^+ + n_-^-}{E}$
\\[3mm]
\hline
\end{tabular}
\caption{Elementary changes of observables $\Delta, \Sigma, \rhop, \rhom, \rhos, \rhou,\rho$ under the four possible flips $(s,p)\to(-s,p)$. The variables $n_s^p$ denote the number of neighbors of class $(s,p)$ of the focal node satisfying $k=\sum_{s,p=\pm}n_s^p$ and $E=\mu N / 2$ is the total number of links.}
\label{tab:elementary_changes}
\end{table}

Putting all ingredients together, Eq.~\eqref{eq:HPA_general} can be evaluated for the variables $\Delta, \Sigma,\rhop, \rhom,\rhos,\rhou$ yielding the expressions
\begin{subequations}
\label{eq:PA_system_app}
\begin{align} 
\frac{d\Delta}{dt}
&=
-\frac{1}{2}
\sum_{s=\pm}\sum_{p=\pm}
s\,x_s^p(1-sp\varepsilon)
\sum_k \frac{P_k}{k}
\left\langle n_{-s}^+ + n_{-s}^- \right\rangle_{s,p,k} ,
\label{eq:ddelta_app}\\
\frac{d\Sigma}{dt}
&=
-\frac{1}{2}
\sum_{s=\pm}\sum_{p=\pm}
sp\,x_s^p(1-sp\varepsilon)
\sum_k \frac{P_k}{k}
\left\langle n_{-s}^+ + n_{-s}^- \right\rangle_{s,p,k},
\label{eq:dsigma_app}\\
\frac{d\rhop}{dt}
&=\frac{1}{\mu}
\sum_k \frac{P_k}{k}
\Big[
x_+^+ (1-\varepsilon)
\left\langle
(n_+^+ - n_-^+)(n_-^+ + n_-^-)
\right\rangle_{+,+,k}+
x_-^+ (1+\varepsilon)
\left\langle
(-n_+^+ + n_-^+)(n_+^+ + n_+^-)
\right\rangle_{-,+,k}
\Big],
\label{eq:drhop_app}\\
\frac{d\rhom}{dt}
&=\frac{1}{\mu}
\sum_k \frac{P_k}{k}
\Big[
x_+^- (1+\varepsilon)
\left\langle
(n_+^- - n_-^-)(n_-^+ + n_-^-)
\right\rangle_{+,-,k}+
x_-^- (1-\varepsilon)
\left\langle
(-n_+^- + n_-^-)(n_+^+ + n_+^-)
\right\rangle_{-,-,k}
\Big],
\label{eq:drhom_app}\\
\frac{d\rhos}{dt}
&=\frac{1}{\mu}
\sum_k \frac{P_k}{k}
\Big[
- x_+^+ (1-\varepsilon)
\left\langle
n_-^- (n_-^+ + n_-^-)
\right\rangle_{+,+,k}
+ x_-^+ (1+\varepsilon)
\left\langle
n_-^- (n_+^+ + n_+^-)
\right\rangle_{-,+,k} \nonumber \\&
\hspace{1.7cm} +
x_+^- (1+\varepsilon)
\left\langle
n_+^+ (n_-^+ + n_-^-)
\right\rangle_{+,-,k}
- x_-^- (1-\varepsilon)
\left\langle
n_+^+ (n_+^+ + n_+^-)
\right\rangle_{-,-,k}
\Big],
\label{eq:drhos_app}\\
\frac{d\rhou}{dt}
&=\frac{1}{\mu}
\sum_k \frac{P_k}{k}
\Big[
x_+^+ (1-\varepsilon)
\left\langle
n_+^- (n_-^+ + n_-^-)
\right\rangle_{+,+,k}
- x_-^+ (1+\varepsilon)
\left\langle
n_+^- (n_+^+ + n_+^-)
\right\rangle_{-,+,k} \nonumber \\&
\hspace{1.7cm}-
x_+^- (1+\varepsilon)
\left\langle
n_-^+ (n_-^+ + n_-^-)
\right\rangle_{+,-,k}
+ x_-^- (1-\varepsilon)
\left\langle
n_-^+ (n_+^+ + n_+^-)
\right\rangle_{-,-,k}
\Big], \label{eq:drhou_app}
\end{align}
\end{subequations}
where $\mu=\sum_k kP_k$ is the average degree of the network and $\langle \cdots \rangle_{s,p,k}$ denotes average values over the multinomial distribution $M_s^p(k,\nb)$, namely 
\begin{align}
\left\langle n_{s'}^{p'} \right\rangle_{s,p,k}&=k\,P(s',p'|s,p), \label{eq:moment1} 
\\ \label{eq:moment2}
\left\langle \left(n_{s'}^{p'}\right)^2 \right\rangle_{s,p,k}
&=
k\,P(s',p'|s,p)
+
k(k-1)\left[P(s',p'|s,p)\right]^2,
\\
\left\langle n_{s'}^{p'} n_{s''}^{p''} \right\rangle_{s,p,k}
&=
k(k-1)\,P(s',p'|s,p)\,P(s'',p''|s,p), \quad\quad (s',p')\neq(s'',p'').
\label{eq:momentcorr}
\end{align}
Using these moments in Eqs.~\eqref{eq:PA_system_app}, and rewriting the result in terms of the variables $\Delta$, $\Sigma$, $\rhop$, $\rhom$, $\rhos$, and $\rhou$, the following closed system of equations is obtained
\begin{subequations}\label{eq:PA_system}
\begin{align}
\frac{d\Delta}{dt}&= \frac{\varepsilon}{2}\left(\rhop-\rhom\right),
\\[2mm]
\frac{d\Sigma}{dt} &=
\frac{1}{2} \left[
\varepsilon(\rhop+\rhom) +(1+\varepsilon)\rhou-(1-\varepsilon)\rhos \right],
\\[2mm]
\frac{d\rhop}{dt}
&=
-\frac\rhop{\mu}+
\frac{\mu-1}{2\mu}
(1-\varepsilon)(\rhop+\rhos)
\left(
\frac{\Sigma+\Delta-\rhop-\rhos}{1+\Delta}
-
\frac\rhop{\Sigma+\Delta}
\right)
\nonumber\\
&\quad+
\frac{\mu-1}{2\mu}
(1+\varepsilon)(\rhop+\rhou)
\left(
\frac{1-\Sigma-\Delta-\rhop-\rhou}{1-\Delta}
-
\frac\rhop{1-\Sigma-\Delta}
\right),
\\[2mm]
\frac{d\rhom}{dt}
&=
-\frac\rhom{\mu}+
\frac{\mu-1}{2\mu}
(1-\varepsilon)(\rhom+\rhos)
\left(
\frac{\Sigma-\Delta-\rhom-\rhos}{1-\Delta}
-
\frac\rhom{\Sigma-\Delta}
\right)
\nonumber\\ &\quad
+\frac{\mu-1}{2\mu}
(1+\varepsilon)(\rhom+\rhou)
\left(
\frac{1-\Sigma+\Delta-\rhom-\rhou}{1+\Delta}
-
\frac\rhom{1-\Sigma+\Delta}
\right),
\\[2mm]
\frac{d\rhos}{dt}
&=
-\frac{1-\varepsilon}{\mu}\rhos
-\frac{(1-\varepsilon)(\mu-1)}{2\mu}\rhos
\left[
\frac{\rhop+\rhos}{\Sigma+\Delta}
+
\frac{\rhom+\rhos}{\Sigma-\Delta}
\right]
\nonumber\\
&\quad+
\frac{(1+\varepsilon)(\mu-1)}{2\mu}
\frac{\Sigma-\Delta}{1-\Delta}
(\rhop+\rhou)
\left[
1-\frac{\rhop+\rhou}{1-\Sigma-\Delta}
\right]
\nonumber\\
&\quad+
\frac{(1+\varepsilon)(\mu-1)}{2\mu}
\frac{\Sigma+\Delta}{1+\Delta}
(\rhom+\rhou)
\left[
1-\frac{\rhom+\rhou}{1-\Sigma+\Delta}
\right],
\\[2mm]
\frac{d\rhou}{dt}
&=
-\frac{1+\varepsilon}{\mu}\rhou
-\frac{(1+\varepsilon)(\mu-1)}{2\mu}\rhou
\left[
\frac{\rhop+\rhou}{1-\Sigma-\Delta}
+
\frac{\rhom+\rhou}{1-\Sigma+\Delta}
\right]
\nonumber\\
&\quad+
\frac{(1-\varepsilon)(\mu-1)}{2\mu}
\frac{1-\Sigma+\Delta}{1+\Delta}
(\rhop+\rhos)
\left[
1-\frac{\rhop+\rhos}{\Sigma+\Delta}
\right]
\nonumber\\
&\quad+
\frac{(1-\varepsilon)(\mu-1)}{2\mu}
\frac{1-\Sigma-\Delta}{1-\Delta}
(\rhom+\rhos)
\left[
1-\frac{\rhom+\rhos}{\Sigma-\Delta}
\right].
\end{align}
\end{subequations}

\section{Reduced pair approximation for the partisan voter model}
\label{app:PA_reduced}

In this Appendix, we derive a reduced pair approximation only in terms of the three variables $\Delta$, $\Sigma$, $\rho$. In contrast to the pair approximation introduced in Sec.~\ref{sec:PA} and detailed in~\ref{app:PA}, here we do not distinguish between the different types of active links according to the preferences of the nodes they connect. As a consequence, the neighborhood composition is characterized solely by the number $n$ of neighbors in the opposite state. We therefore describe the dynamical evolution of a variable $X$ as
\begin{align} \label{eq:HPA_general_reduced_app}
 \frac{dX}{dt}\approx&\frac{1}{\delta t}\sum_{k}\sum_{s=\pm}\sum_{p=\pm}\sum_{n=0}^kP(s,p,k,n) \times P_{s\to-s}(s,p,k,n)\delta X(s\to-s, n),
\end{align}
where $\delta t=1/N$, $P(s,p,k,n)$ is the probability of selecting a node of class $(s,p)$, degree $k$, and $n$ neighbors in opposite state. The term $P_{s\to -s}(s,p,k,n)$ is the probability that this node flips its state, given by
\begin{equation}
P_{s\to -s}(s,p,k,n) = \frac{n}{k} \left(\frac{1-s\, p\,\varepsilon}{2}\right).
\label{eq:p_flip_app}
\end{equation}
The elementary changes $\delta X(s\to-s, n)$ of the variables $\Delta, \Sigma,\rho$ under a flip $(s,p)\to(-s,p)$ are shown in Table~\ref{tab:elementary_changes}. For $\rho$ they can be written solely in terms of the number $n$ of neighbors in opposite state as
\begin{align} \label{eq:deltarho_reduced_app}
\delta\rho(s\to -s,n) &= \frac{k-2n}{E},
\end{align}
where $E=\mu N/2$ is the total number of links.

Under this reduced pair approximation, we assume that the active links are distributed according to the abundance of each node class such that the number of active links incident to nodes of class $(s,p)$ is $N\mu\rho x^p_s/2x_s$. Thus, the conditional probability of selecting a node in opposite state $P(-s|s,p)$ is approximated by
\begin{equation}
P(-s|s,p)\approx P(-s|s)=\frac\rho{2x_s},
\label{eq:conditional_state_reduced}
\end{equation}
independently of the preference $p$.

Therefore, the number $n$ of neighbors in the opposite state of a node of class $(s,p)$ and degree $k$ follows a binomial distribution,
\begin{equation}
B_s(k,n)
=
\binom{k}{n}
\left(\frac\rho{2x_s}\right)^{n}
\left(1-\frac\rho{2x_s}\right)^{k-n}.
\label{eq:binomial_reduced_app}
\end{equation}
Accordingly, the probability $P(s,p,k,n)$ of selecting a node of class $(s,p)$, degree $k$, and with $n$ neighbors in opposite state is
\begin{equation}
P(s,p,k,n)=x_s^p\,P_k\,B_s(k,n),
\label{eq:P_spkn_reduced_app}
\end{equation}
where $P_k$ is the degree distribution.

Using Eq.~\eqref{eq:HPA_general} together with Eqs.~\eqref{eq:deltarho_reduced_app}--\eqref{eq:binomial_reduced_app}, one obtains
\begin{align}
\frac{d\Delta}{dt}
&=
-\frac{1}{2}
\sum_{s=\pm}\sum_{p=\pm}
s\,x_s^p(1-sp\varepsilon)
\sum_k \frac{P_k}{k}
\left\langle n\right\rangle_{s,p,k},
\label{eq:dDelta_reduced_app}
\\
\frac{d\Sigma}{dt}
&=
-\frac{1}{2}
\sum_{s=\pm}\sum_{p=\pm}
sp\,x_s^p(1-sp\varepsilon)
\sum_k \frac{P_k}{k}
\left\langle n\right\rangle_{s,p,k},
\label{eq:dSigma_reduced_app}
\\
\frac{d\rho}{dt}
&=
\frac{1}{
\mu}
\sum_{s=\pm}\sum_{p=\pm}
x_s^p(1-sp\varepsilon)
\sum_k \frac{P_k}{k}
\left\langle
\bigl(k-2n\bigr)n
\right\rangle_{s,p,k},
\label{eq:drho_reduced_app}
\end{align}
where $\mu=\sum_k kP_k$ is the average degree of the network, and $\langle\cdots\rangle_{s,p,k}$ denotes the average over the binomial distribution in Eq.~\eqref{eq:binomial_reduced_app}. Using the first two moments of the binomial distribution,
\begin{align}
\left\langle n\right\rangle_{s,p,k}
&=
k\,\frac\rho{2x_s},
\\
\left\langle n^2\right\rangle_{s,p,k}
&=
k\,\frac\rho{2x_s}
+
k(k-1)\left(\frac\rho{2x_s}\right)^2,
\end{align}
and rewriting the result in terms of the variables $\Delta$, $\Sigma$, and $\rho$, we obtain
\begin{subequations}
\label{eq:PA_reduced_system}
\begin{align} 
\frac{d\Delta}{dt}
&=\frac{\varepsilon \Delta \rho (1-2 \Sigma ) }{1-4 \Delta ^2} 
\\
\frac{d\Sigma}{dt}
&=\frac{1}{2}\frac{\rho \left( 1 + \varepsilon-4 \varepsilon \Delta ^2 -2 \Sigma \right)}{1-4 \Delta ^2}
\\
\frac{d\rho}{dt}
&= \frac\rho{\mu} 
\Bigg[
\frac{1+2\Delta-\varepsilon(2\Sigma-1)}{1+2\Delta}
\left(
\mu-2-\frac{2(\mu-1)\rho}{1+2\Delta}
\right)+
\frac{1-2\Delta-\varepsilon(2\Sigma-1)}{1-2\Delta}
\left(
\mu-2-\frac{2(\mu-1)\rho}{1-2\Delta}
\right)
\Bigg].
\end{align}
\end{subequations}

This dynamical system presents:
\begin{itemize}
 \item Two non-hyperbolic saddle solutions 
 \begin{equation}
 (\Delta_\st,\Sigma_\st,\rho_\st)=\left(\pm \frac{1}{2},\,\frac{1}{2},\,0\right),
 \end{equation}
 corresponding to the two absorbing consensus states.
 \item One stable coexistence solution, \begin{equation} (\Delta_\st,\Sigma_\st,\rho_\st)=\left(0,\,\frac{1+\varepsilon}{2},\,\frac{\mu-2}{2(\mu-1)}\right).
 \end{equation}
\end{itemize}
Importantly, the reduced pair approximation yields the same set of stationary solutions as the extended pair approximation. Hence, although the reduced closure neglects the distinction among the different types of active links, it still captures the global stationary density of active links. What is lost in the reduced description is not the stationary structure itself, but the information on how the total density of active links is distributed among the different link classes, together with an accurate description of the transient dynamics.
\end{document}